\documentclass[twocolumn]{aastex631}
\usepackage{amsmath}

\newcommand{\Rback}{R}

\newcommand{\eqb}{\begin{eqnarray}}
\newcommand{\eqe}{\end{eqnarray}}
\newcommand{\eqbn}{\begin{eqnarray*}}
\newcommand{\eqen}{\end{eqnarray*}}

\newcommand{\diff}{{\rm d}}

\begin{document}

\title{Angular Power Spectrum of TeV-PeV Cosmic Ray Anisotropies}

\author{Wenyi Bian}
\email{bianwenyi@sjtu.edu.cn}
\affiliation{%
 Tsung-Dao Lee Institute, Shanghai Jiao Tong University, Shanghai 201210, P. R. China
}
\affiliation{School of Physics and Astronomy, Shanghai Jiao Tong University, Shanghai 200240, P. R. China}

\author{Gwenael Giacinti}%
\email{gwenael.giacinti@sjtu.edu.cn}
\affiliation{%
 Tsung-Dao Lee Institute, Shanghai Jiao Tong University, Shanghai 201210, P. R. China
}
\affiliation{School of Physics and Astronomy, Shanghai Jiao Tong University, Shanghai 200240, P. R. China}
 
\author{Brian Reville}
\affiliation{Max-Planck-Institut f\"ur Kernphysik, Postfach 103980, 69029 Heidelberg, Germany}



\begin{abstract}
Simulations of the cosmic-ray (CR) anisotropy 
down to TeV energies are presented, using turbulence parameters consistent with those inferred from observations of the interstellar medium.
We compute the angular power spectra $C_{\ell}$ of the CR anisotropy obtained from the simulations.
We demonstrate that the amplitude of the large scale gradient in the CR density profile
affects only the overall normalisation of the $C_{\ell}$s,
without affecting the shape of the angular power spectrum.
We show that the power spectrum depends on CR energy, and that it is sensitive to the location of the observer at small $\ell$.
It is found to flatten at large $\ell$, and can be modelled by a broken power-law, exhibiting a break at $\ell \approx 4$.
Our computed power spectrum at $\sim 10\,$TeV fits well HAWC and IceCube measurements.
Moreover, we calculate all coefficients of the spherical harmonics 
and compute the component of the angular power spectrum projected onto the direction of the local magnetic field line.
We find that deviations from gyrotropy become increasingly important at higher CR energies and larger values of $\ell$.
\end{abstract}


\section{\label{sec1:1}Introduction}

The propagation of Galactic cosmic rays (CRs) within the galaxy follows a diffusion process, where the turbulent magnetic fields in the interstellar medium (ISM) randomize the momentum of CRs over a sufficiently long period of propagation. Consequently, the arrival directions of Galactic CRs are close to isotropic.
Existing experiments however measure a small anisotropy in the CR flux observed at Earth. In this article, we focus on the anisotropy of TeV to PeV Galactic CRs. Anisotropy in this energy range has been measured by several experiments, with an amplitude of the order of $10^{-3}$~ \citep{di2014observation}. This implies the existence of a weak spatial gradient of the CR density in the ISM close to the Earth. 
This gradient is thought to be created by either recent CR sources near the Earth, or a gradient in the source density in the Galactic disc, or CR escape in the halo~\citep{blasi2012diffusive}.
A few studies have attempted to explain the amplitude and phase variations of the TeV-PeV dipole moment anisotropy in the observed data, see e.g. \cite{ahlers2016deciphering}.
The large-scale structures of the anisotropy, especially features related to the dipole moment, have received most of the attention \citep{blasi2012diffusive,pohl2013understanding,kumar2014large,mertsch2015solution}, despite measurements of small-scale anisotropy structures by e.g. ARGO-YBJ \citep{bartoli2013medium}, Tibet AS-$\gamma$ \citep{amenomori2006anisotropy}, Milagro 
\citep{abdo2008discovery}, IceCube \citep{aartsen2016anisotropy}, and HAWC \citep{abeysekara2014observation}.  The small-scale anisotropy structures are a consequence of the turbulent magnetic field permeating the interstellar medium \citep{giacinti2012local,AhlersPRL2014,battaner2015magnetic,lopez2016cosmic}.

The angular power spectrum can be used to describe the level of anisotropy at different scales and directions. It has been employed in experiments studying the CR anisotropy, providing a valuable means to characterize the directional distribution of CRs across various angular scales.
A combined all-sky anisotropy map at $10$\,TeV has been presented by the HAWC and IceCube collaborations in \cite{abeysekara2019all},
where the angular power spectrum is also presented, though saturated by noise above $\ell\approx14$. 

Previous works have provided both analytical~\citep{AhlersPRL2014} and numerical~\citep{ahlers2015small,lopez2016cosmic,kumar2019} calculations of the angular power spectrum of small-scale anisotropies. Such numerical calculations have been made by propagating individual CRs either in synthetic turbulence~\citep{ahlers2015small,kumar2019} or in 3D MHD simulations~\citep{lopez2016cosmic}. Due to the finite size of MHD grids, \cite{lopez2016cosmic} restricted their calculations to CR energies larger than 750\,TeV. In principle, simulations with synthetic turbulence do not have such a constraint on CR energies. However, the computing time becomes increasingly large at smaller energies, because the particle gyration radius is proportional to its energy. In \cite{ahlers2015small} and \cite{kumar2019}, the particle gyroradius was set to one tenth of the coherence length of the turbulence, which corresponds to CR energies $\gtrsim 1$\,PeV for typical interstellar turbulence parameters ---a few $\mu$G magnetic field strength and $\sim 10$\,pc coherence length. In contrast, most measurements of the power spectrum of the CR anisotropy exist for $\sim (1-10)$\,TeV CRs, where more statistics are available~\citep{abeysekara2019all}. This corresponds to CR gyroradii smaller by 2 to 3 orders of magnitude than those in any existing simulations. At TeV energies, substantially less power is expected to be present in the modes that scatter CRs, and, at any point in space, magnetic field lines should also look substantially straighter and more ordered on the gyroradius scale, than at PeV energies. Therefore, the power spectrum may be different at TeV energies, as confirmed below.

In the present paper, we provide the first numerical simulations of the CR anisotropy down to TeV energies. We calculate the anisotropy of CRs in the TeV-PeV energy range, which is the relevant range for the interpretation of available experimental data. By introducing a large-scale CR gradient 
across a turbulent magnetic field and backtracking the CRs, we obtain our anisotropy results.  In section\,\ref{method}, we provide an overview of the numerical simulation method, the configuration of the turbulent magnetic field, the procedure for computing angular power spectra, the fitting of the noise in the angular power spectra, the impact of CR backtracking distances on the angular power spectra, and the simulation results within the TeV-PeV energy range. In section\,\ref{Energy}, we show the dependence of anisotropy structures on energies and observer locations. By employing a broken power-law fit, we demonstrate that the dependence of large-scale structures and small-scale structures on energy and location differs. 
In section\,\ref{Rotate}, we analyze the expansion coefficients of anisotropy structures symmetric along the local magnetic field direction or dipole direction. This reveals that large- and medium-scale anisotropy structures are more gyrotropic than small-scale anisotropies. In section\,\ref{Discuss&Conclusion}, we discuss our results and conclude.

\begin{figure*}
 \includegraphics[width=\textwidth]{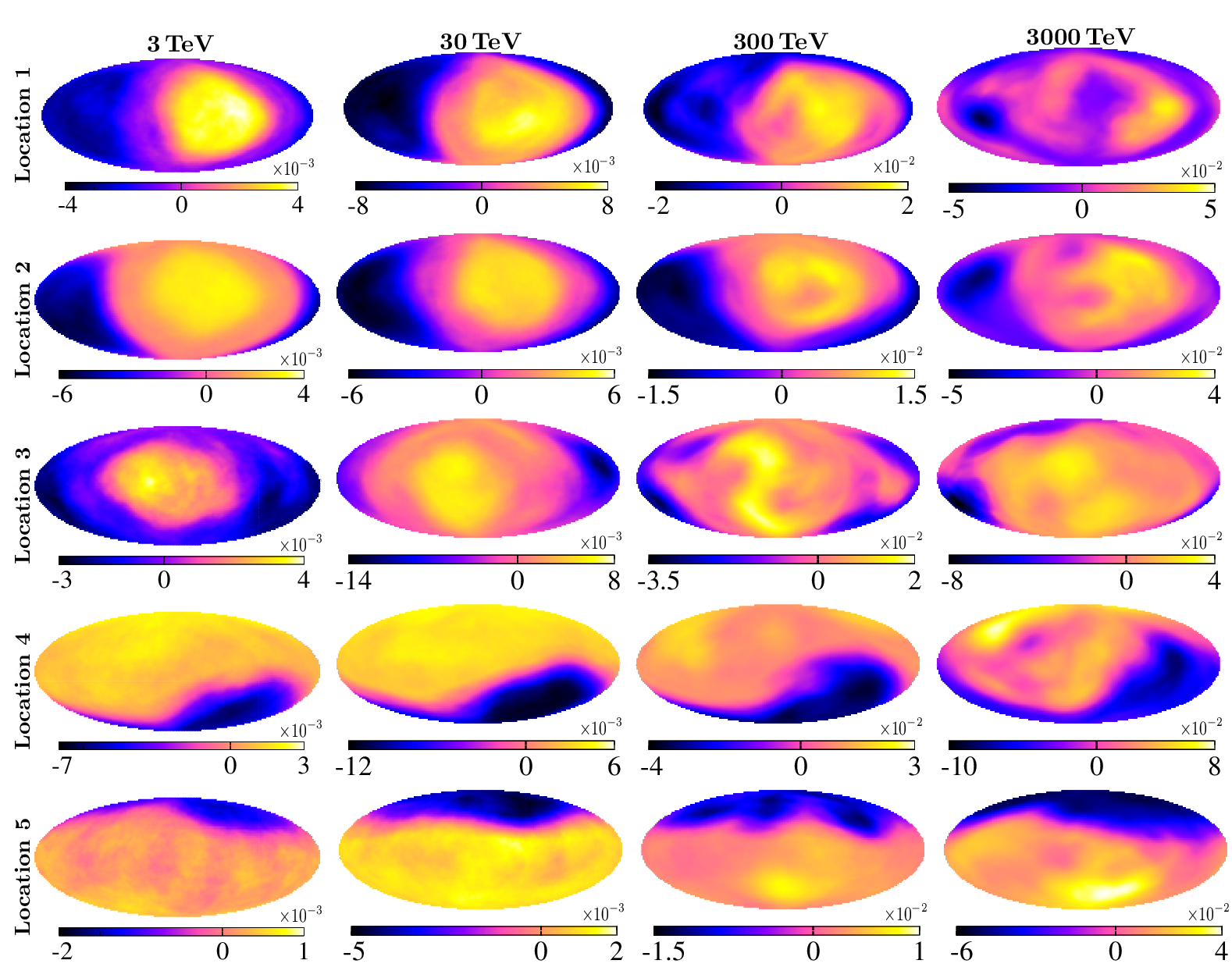}	
    \caption{Anisotropy sky-maps from the simulations.
    Each row's panels represent the results at one same observer location with energy ranging from 3$\,$TeV to 3000$\,$TeV,
    while each column's panels show the results at different observer locations with the same CR energy.
    The number of CR particles in 19 simulations has been set to 1 million. 
    For the simulation at Location\,1 and with the energy of 3 TeV, 
    the number of particles in the simulation is only 0.5 million.
    In all 20 simulation results, the radius of the backtracking surface is 150$\,$pc. 
    All the anisotropy sky maps are smoothed over a 20 degree radius.}
    \label{fig:all_simulation_skymap}
\end{figure*}

\section{\label{method}Numerical Method}
\subsection{Numerical simulation method}

In this section, we introduce our method for calculating CR anisotropies and their angular power spectra.
The numerical method matches closely the work of \cite{giacinti2012local}.
We select the observer's location within the turbulent magnetic field 
and backtrack CR particles with random initial directions from that point.
The backtracked CRs will reach a spherical surface 
whose center is the observer location and radius is denoted by $\Rback$.
The value of $\Rback$ should be bigger than the scattering length of the CRs 
to ensure that the propagation of CRs can be considered as a diffusive process.
In the following section, 
we will refer to this spherical surface as the backtracking surface.
With the positions of backtracked CR particles on the backtracking surface and 
the weights for each particle's backtracking path, 
we obtain the anisotropy results of the simulations as in \cite{giacinti2012local}.
We constructed the 3D synthetic turbulent magnetic field by using the \lq\lq nested grids\rq\rq\/ method of \cite{giacinti2012cosmic}. 
The construction of the entire magnetic field 
involves calculating the magnetic field in the whole space 
and storing the turbulent magnetic field in nested grids, which contain three different sizes and where 
grids with different sizes are repetitively arranged to cover the entire space.
A turbulent magnetic field with Kolmogorov power spectrum ($P(k)\propto k^{-\frac{5}{3}}$) is used. 
AMS-02 measurements of the Boron-to-Carbon ratio are compatible with Kolmogorov turbulence~\citep{AMS_BoC}, at least up to $\sim 2$\,TV CR rigidity.
The three different sizes of the repeated boxes are such that the spacings between their grid points are: 
$\Delta_1=1.25\,\mathrm{pc}, \Delta_2=2.5\times10^{-2}\,\mathrm{pc}$ and $\Delta_3=5\times10^{-4}\,\mathrm{pc}$.
The number of vertices per grid is $\mathrm{N}^3=256^3$.
The minimum scale in the magnetic field for the biggest box is set to $l_{\rm min,1}=2.5\,$pc,
and the ratio between the maximum and minimum scales is ${l_{\rm max,1}}/{l_{\rm min,1}}=60$.
This corresponds to an outer scale ${l_{\rm max,1}}=150\,$pc for the turbulence, and to a coherence length of $30\,$pc, which is typical of ISM values.
For the boxes of intermediate size, $l_{\rm min,2}$ is set to $5\times10^{-2}\,\mathrm{pc}$ and the ratio is ${l_{\rm max,2}}/{l_{\rm min,2}}=50$.
For the smallest boxes, $l_{\rm min,3}$ is set to $10^{-3}\,\mathrm{pc}$ and the ratio is ${l_{\rm max,3}}/{l_{\rm min,3}}=50$.
In Figure~\ref{fig:all_simulation_skymap}, we show the skymaps for five randomly selected positions (denoted hereafter Location 1-5)
within the turbulent magnetic field and 
calculate the anisotropy results for different CR energies. 

\begin{figure}
 \includegraphics[width=0.45\textwidth]{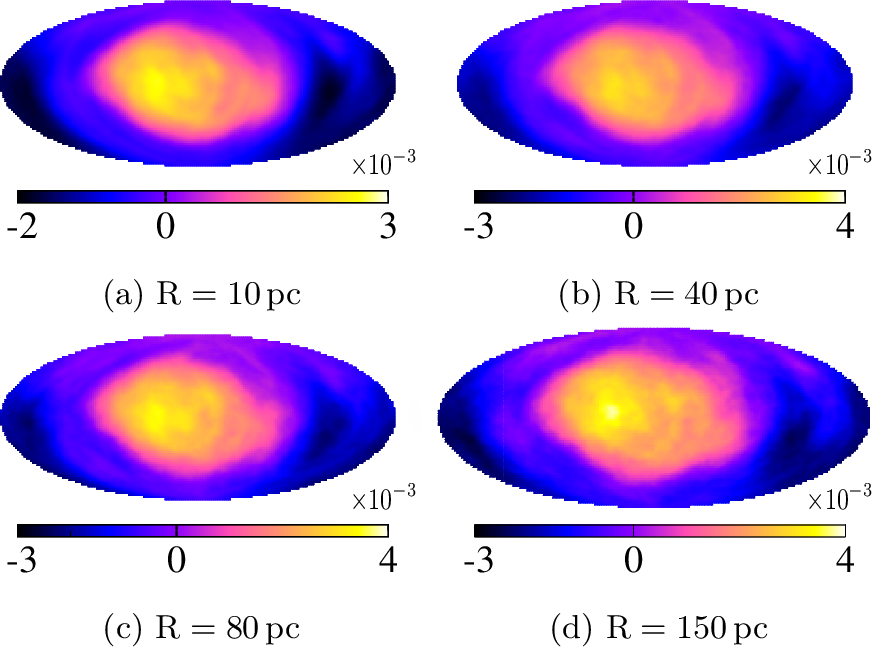}	
  \caption{Anisotropy sky-maps at different backtracking surfaces. The skymap is for the observer Location\,3 and 3$\,$TeV. All the skymaps look similar. The overall amplitude slightly increases with the radius increasing.}
  \label{fig:diff_R_skymap}
\end{figure}

\begin{figure}
  \centering
    \includegraphics[width=0.9\linewidth]{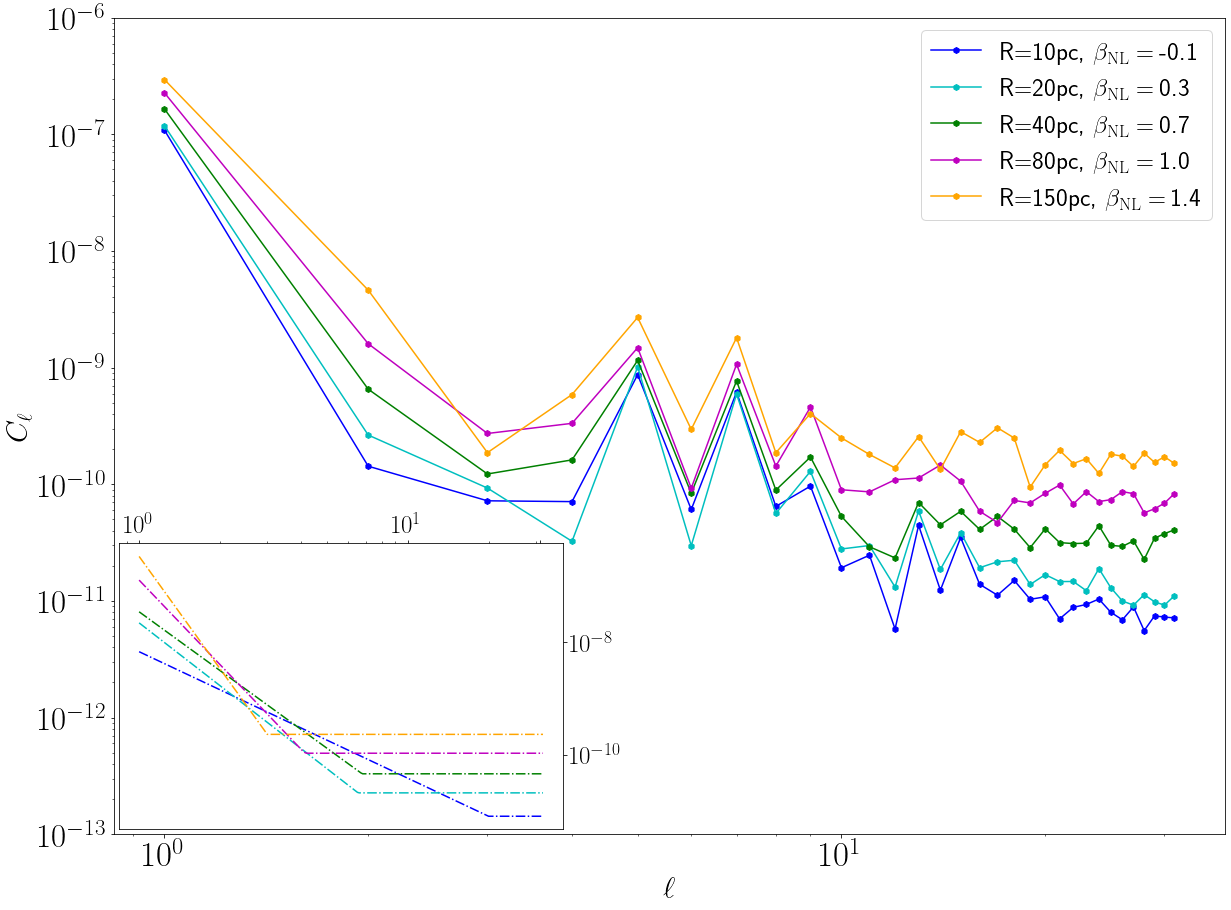}
    \includegraphics[width=0.9\linewidth]{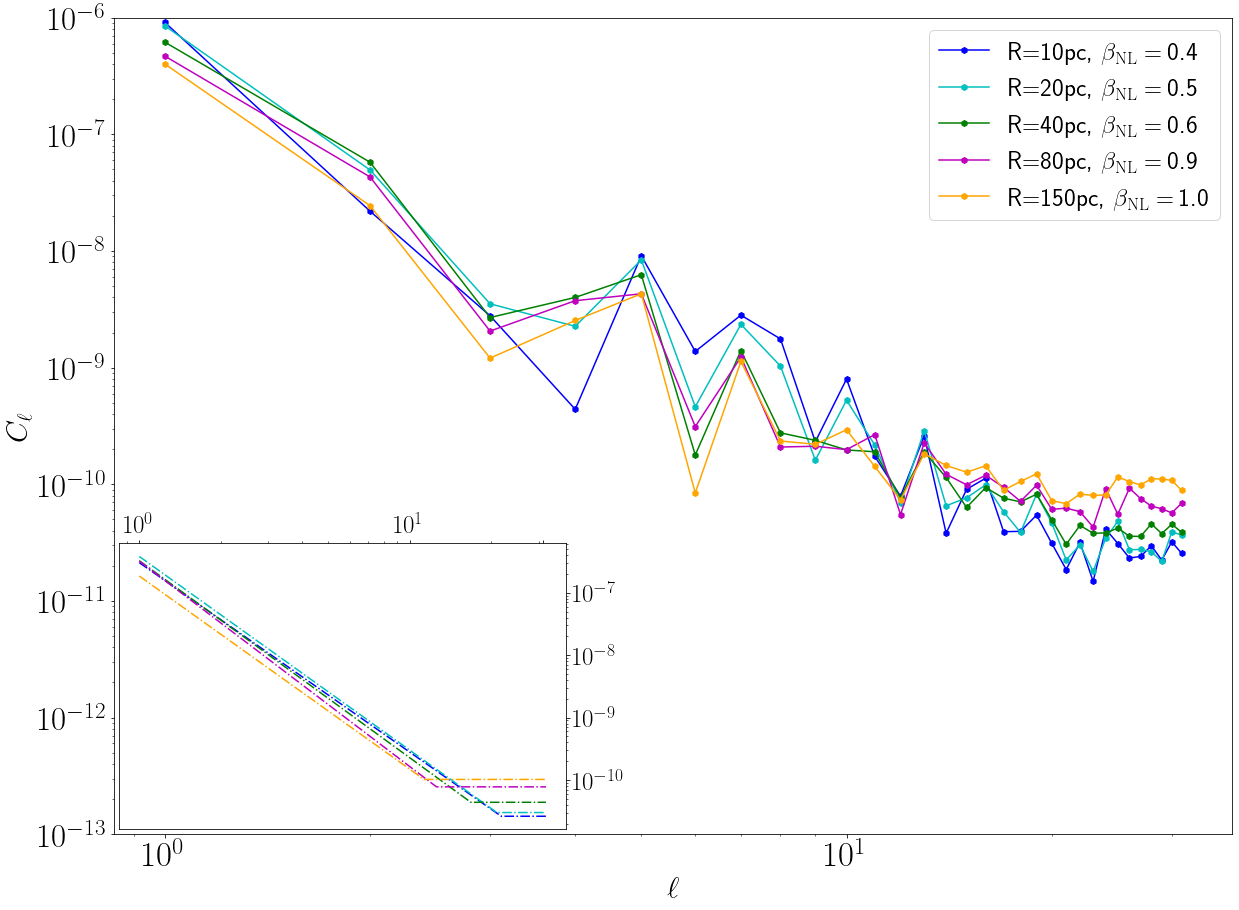}
    \includegraphics[width=0.9\linewidth]{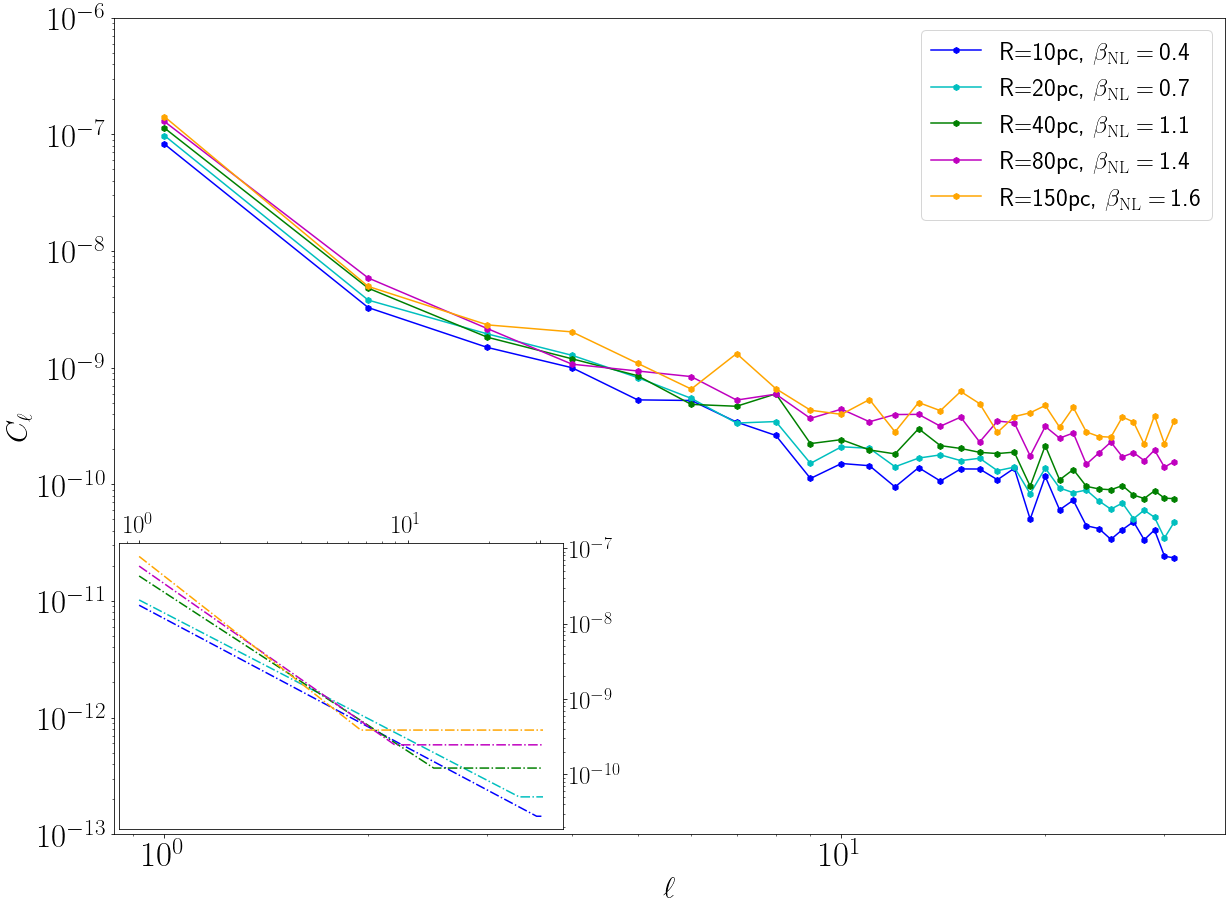}
  \caption{Angular power spectrum for different backtracking radii at different observer locations. The three panels from top to bottom correspond to Location\,1-3. 
  The inset in the left bottom corner shows our broken power-law fit that takes into account the noise level at large $\ell$ (flat part).
  The CR energy in these three panels is fixed at 3\,TeV.}
  \label{fig:diff_R_spectrum}
\end{figure}

\subsection{Calculating the CR anisotropy angular power spectrum}
For each test particle in the simulation, we record its \lq\lq arrival\rq\rq\/ direction at the observer location and the corresponding flux.
We divide the entire skymap into $2\times {L_{\rm max}}^2$ bins (${L_{\rm max}}$ bins along the $\theta$-direction and $2{L_{\rm max}}$ bins along the $\phi$-direction.) The grid points are uniformly distributed in the $\phi$-direction. The sampling method for grid points in the $\theta$-direction uses the Gauss-Legendre quadrature points, the details are shown in Appendix \ref{sec:appendixA} and \cite{NumericalRecipes}. We calculate the average CR flux $f(\vec{n_\alpha})$ (the ratio of the total flux which sums up all test CR flux inside the angular bin and the total number of CRs inside this bin, and $\vec{n_\alpha}$ represent the vector for the angular bin $\alpha$). The spherical harmonic expansion of  the CR anisotropy at a fixed energy, $f(E,\theta,\phi)$, is as follows: 
\begin{equation}
  f(E,\theta ,\phi )=\sum_{\ell=0}^{L_{\max}}\sum_{m=-\ell}^{\ell}f_{\ell}^m(E)Y_{\ell}^m(\theta ,\phi) \, .
\end{equation}
All the coefficients $f_l^m (E)$ will be abbreviated as $f_l^m$.
Then the angular power spectrum for the anisotropy is defined as in \cite{abeysekara2019all}, and each anisotropy result within each angular bin is re-normalized according to the same reference:
\begin{align}
  &C_{\ell}=\frac{1}{2\ell+1}\sum_{m=-\ell}^{\ell}\left | f_{\ell}^m \right |^2\, ,\\
  &\delta I_{\alpha }=\frac{N_{\alpha }-\left \langle N_{\alpha } \right \rangle }{\left \langle N_{\alpha } \right \rangle}    \, . 
\end{align} 
Here, $N_{\alpha }$ is the average CR flux in the angular bin $\alpha$, and $\left \langle N_{\alpha } \right \rangle$ is the average value of all angular bins.
In the simulations, due to the finite number of CRs, the angular power spectrum gradually approaches the noise level as $\ell$ increases.
The noise in the spectrum manifests itself as fluctuations in the $C_{\ell}$ values around a constant when $\ell$ exceeds a certain value.
When calculating the angular power spectrum, 
we use a constant function form for the power spectrum when the value of the variable $\ell$ is bigger than a 
free parameter $\ell_{\rm eff}$. 
We designate the noise level as $\mathcal{N}$ and define the noise level parameter $\beta_{\mathrm{NL}}$ using the following formula:
\begin{eqnarray}
  \beta_{\mathrm{NL}}=\log_{10}\left(\frac{\mathcal{N}}{10^{-11}}\right) \, .
  \label{eq:beta_index}
\end{eqnarray} 
In general, 
as the backtracking time for CRs in the simulation increases, 
the angular power spectrum falls below the noise level \citep{ahlers2015small}.
At $\sim 100$\,TeV, propagating 1 million CRs is sufficient to ensure that the noise level in our simulations does not affect the multipoles up to $\ell = 32$.
However, 
when the CR energy is of the order of several TeV, 
the angular power spectrum reaches the noise level at $\ell<32$.
In order to mitigate the impact of noise in the angular power spectra 
at energies of the order of several TeV, 
the number of particles required in the simulation needs to be greatly increased, 
by several orders of magnitude. 
Alternatively, we can decrease the backtracking time of CRs by reducing the radius $\Rback$ of the backtracking surface. 
We show in the following that this is a highly effective method for reducing noise in the simulation results.

\begin{figure}
    \centering
    \includegraphics[width=0.9\linewidth]{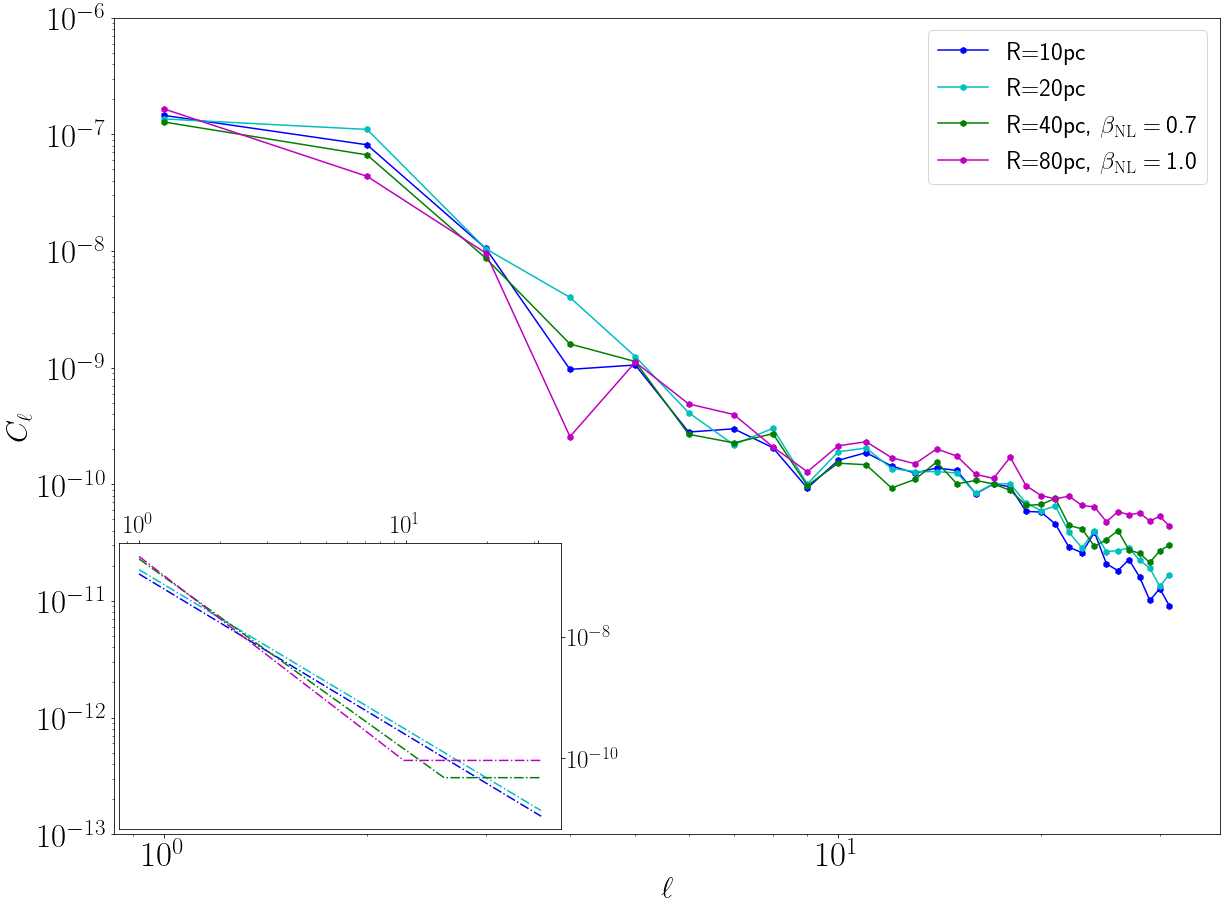}    
    \includegraphics[width=0.9\linewidth]{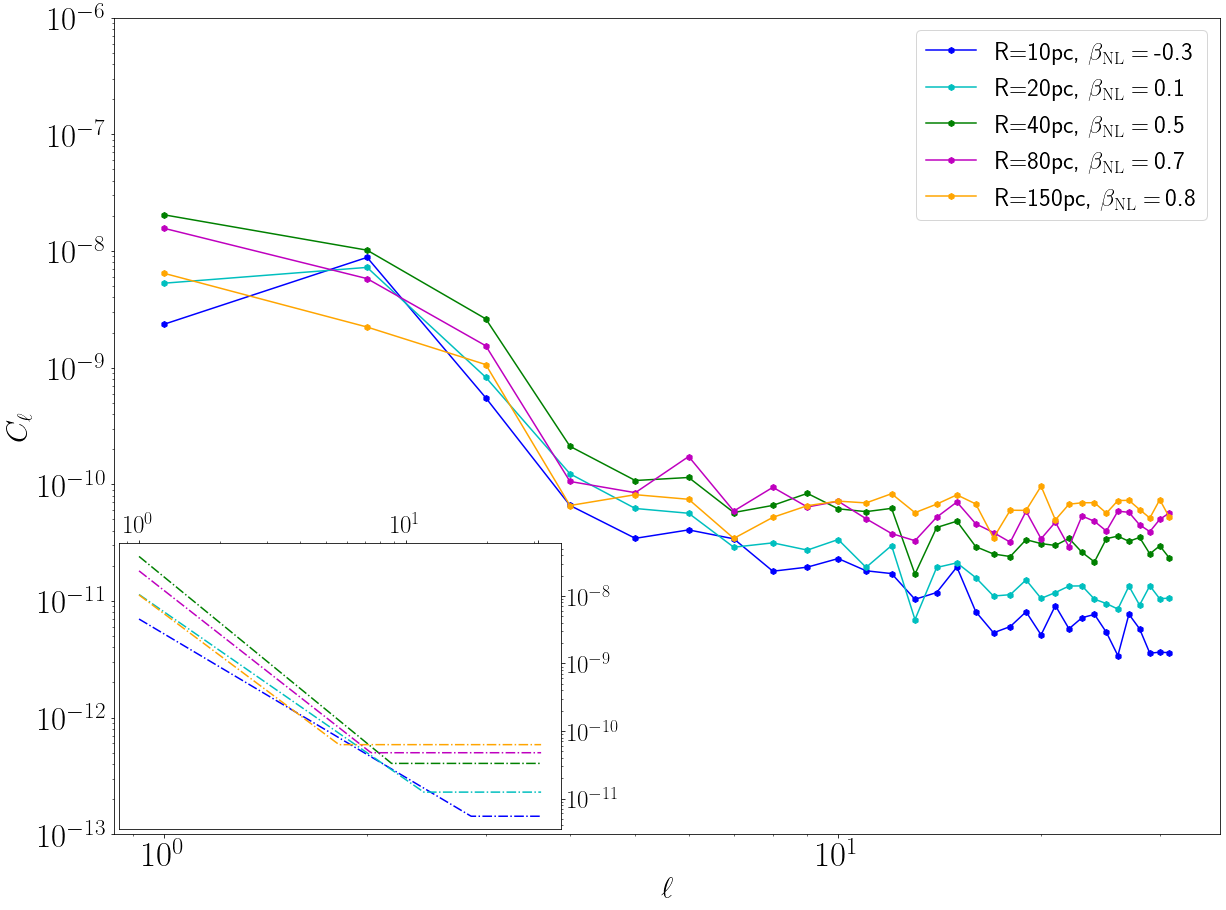}    
  \caption{Same as in Figure~\ref{fig:diff_R_spectrum}, but for an observer at Location\,4 (upper panel) and Location\,5 (lower panel).}
  \label{fig:diff_R_spectrum_2}
\end{figure}

\subsection{Angular power spectrum at 3$\,$TeV with a varying backtracking radius}
In Figure~\ref{fig:diff_R_skymap}, 
we demonstrate the impact of changing the radius of the backtracking surface 
at Location 3 on the anisotropy results.
The radius of the backtracking surface was gradually reduced from the initial setting of 150$\,$pc to 10$\,$pc 
in the simulations.
Apart from a slight reduction in the overall amplitude of the anisotropy, 
there are no significant differences in the structure of the anisotropy sky maps.
In Figures~\ref{fig:diff_R_spectrum} and \ref{fig:diff_R_spectrum_2}, 
we display the angular power spectrum with $L_{\rm max}$ set to 32 for 3$\,$TeV energy 
at all the five positions. 
We fit the noise level in the angular power spectrum using the following function form:
\begin{eqnarray}
  C_{\ell} = \begin{cases}
    \mathcal{N} \left(\frac{\ell}{\ell_{\text{eff}}}\right)^{\alpha}, & \text{if } \ell \le \ell_{\text{eff}}, \\
    \mathcal{N}, & \text{if } \ell > \ell_{\text{eff}}.
  \end{cases}
  \label{eq:noise_level}
\end{eqnarray} 
In the fits, $\ell_{\text{eff}}$ is a free parameter.
If $\ell_{\text{eff}}$ is bigger than 32, we consider that these spectra do not have any noise.
In these two Figures, 
we show the fits of each spectra in the left bottom subfigure in each panel.
It can be observed 
that the noise level in the spectrum at all five locations decreases
as the value of $\Rback$ is reduced, and the values of $\ell_{\text{eff}}$ become larger.
Furthermore, at most positions, reducing the value of $\Rback$  
does not have a significant impact on the power-law fitting of the angular power spectrum at $\ell < \ell_{\text{eff}}$.
Only at Location 1, 
the parameter $\alpha$ for the power-law fit at $\ell < \ell_{\text{eff}}$
shows more noticeable variations.
This is because at this location, 
the weights of multipoles with $\ell$ equal to 5 and 7 are significantly greater 
than the weights of multipole structures with $\ell$ equal to 4, 6 and 8. 
Due to this, the fitting results reach the noise level more quickly.
This also suggests 
that it might be more reasonable to separately fit the angular power spectra for large-scale anisotropy and small-scale anisotropy structures. 
This will be discussed in Section~\ref{Energy}.
In the following sections, 
for all simulations at 3$\,$TeV, 
we set the value of $\Rback$ to 10$\,$pc.
For all the other energies, we set $\Rback=150\,$pc.

\begin{figure}[t]
  \centering
  \includegraphics[width=0.9\columnwidth]{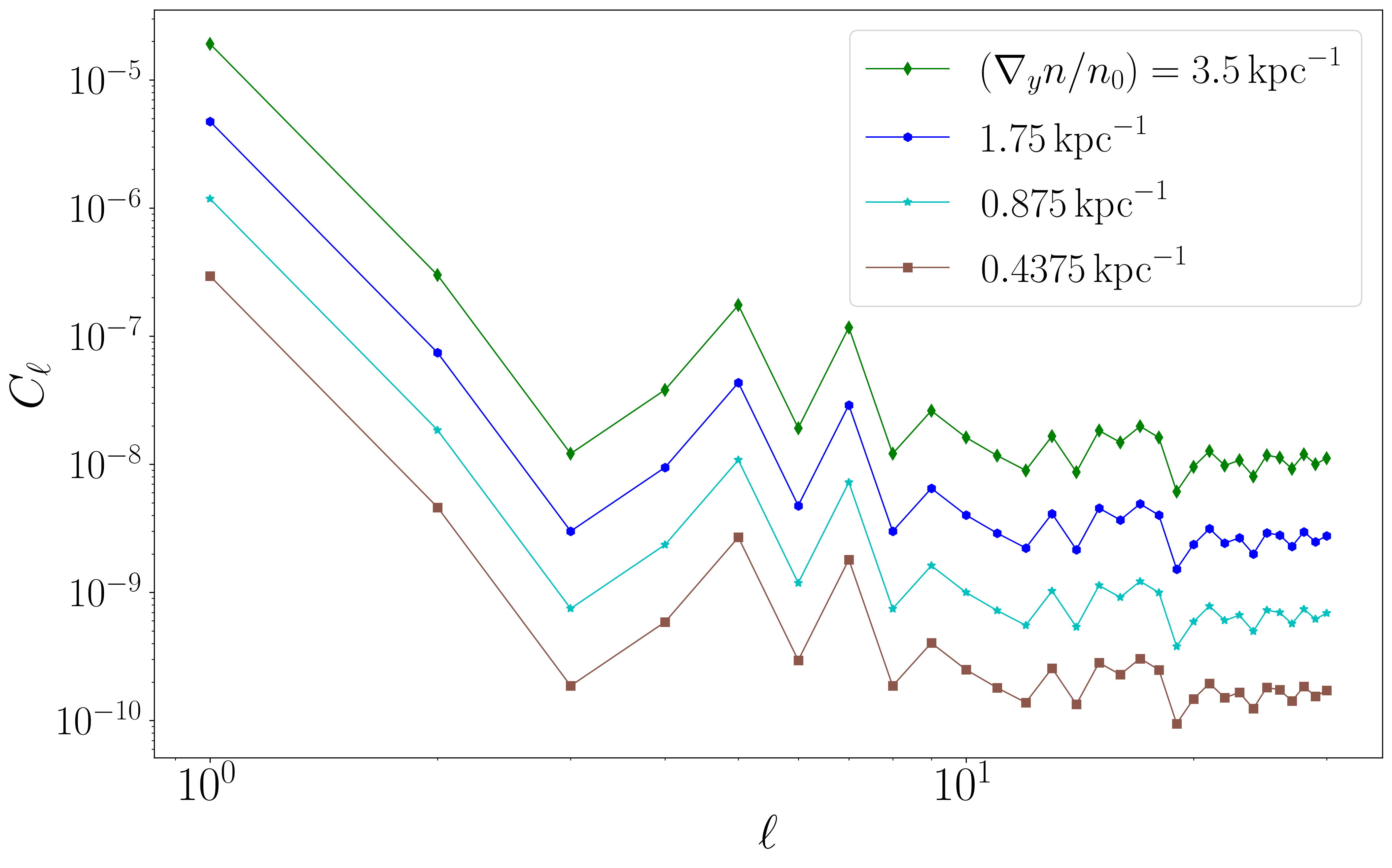}
  \caption{Angular power spectra at Location 2 at 3\,TeV CR energy, calculated for four different CR density gradients. The initial value of ${\nabla n/n_0}$ in the y-direction is set to 3.5$\,\mathrm{kpc}^{-1}$ and  the other three are divided by powers of 2.
  }
  \label{fig:grad}
\end{figure}

\begin{figure}[t]
  \centering
    \includegraphics[width=\linewidth]{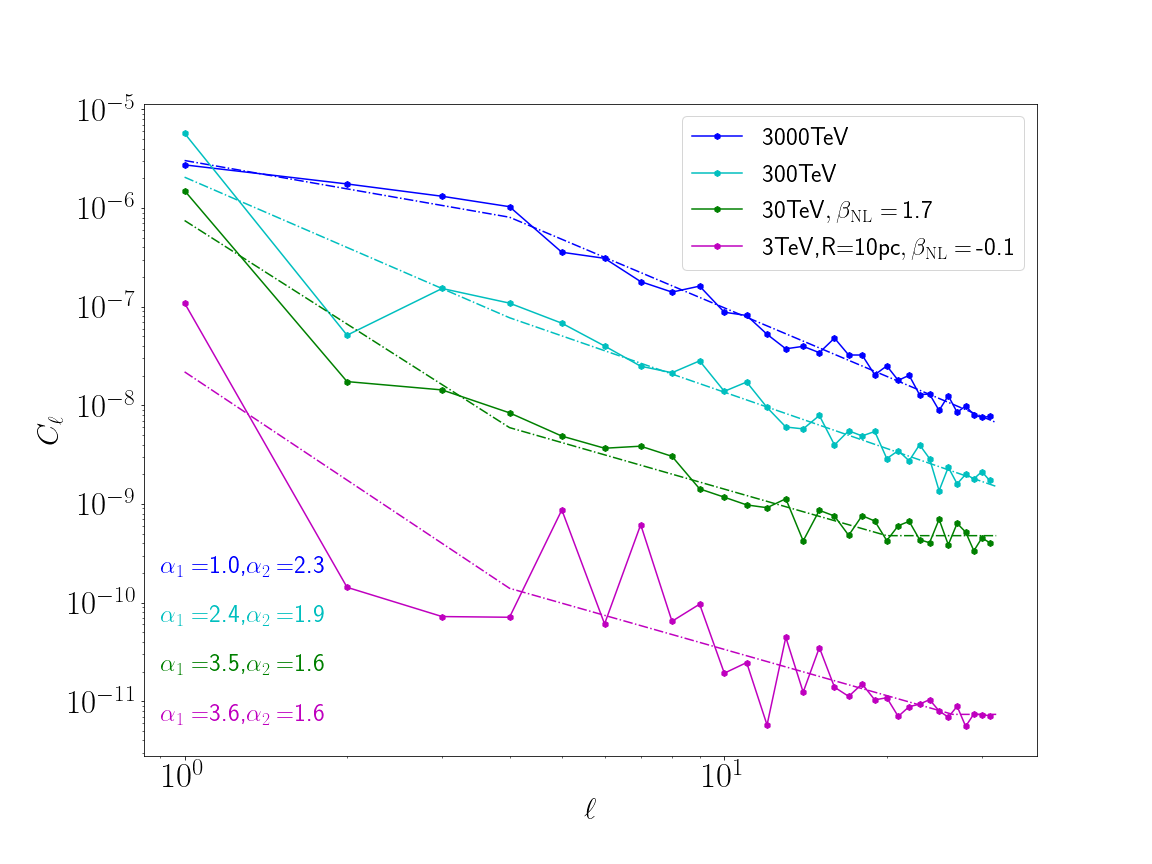}
    \includegraphics[width=\linewidth]{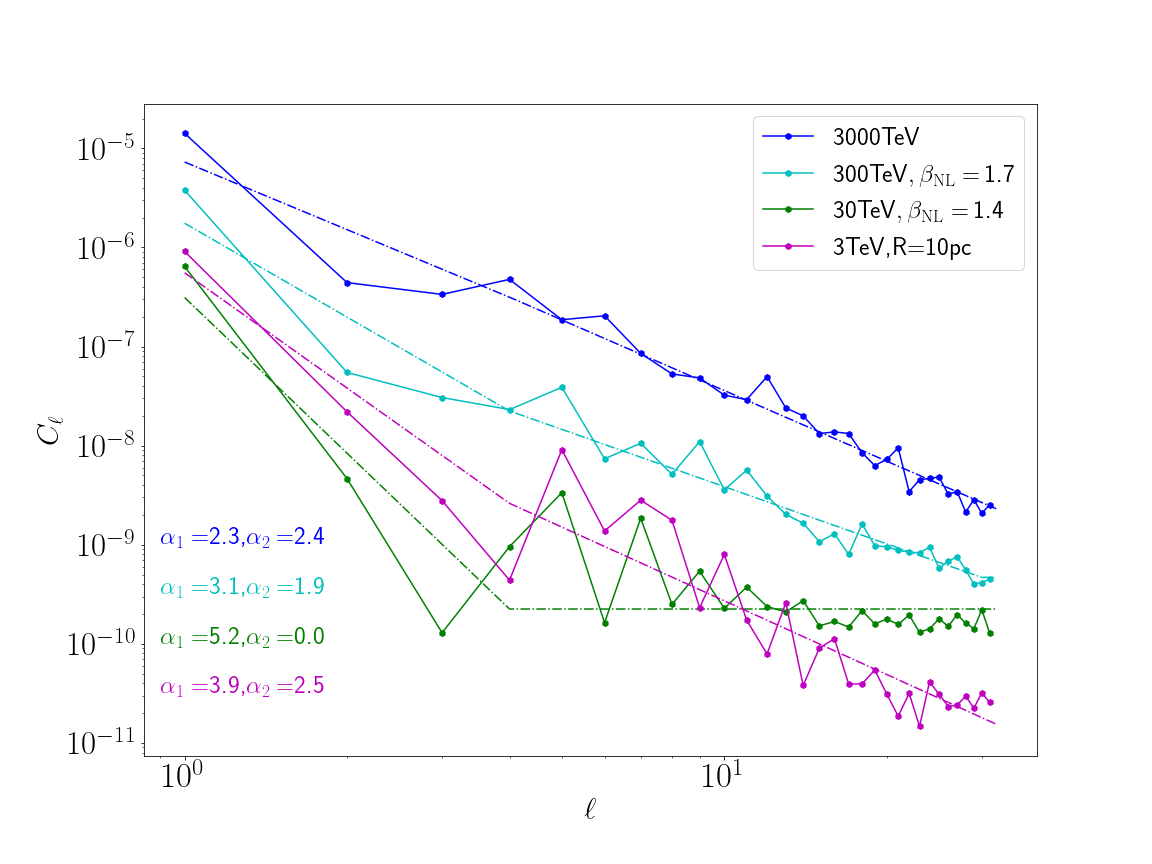}
    \includegraphics[width=\linewidth]{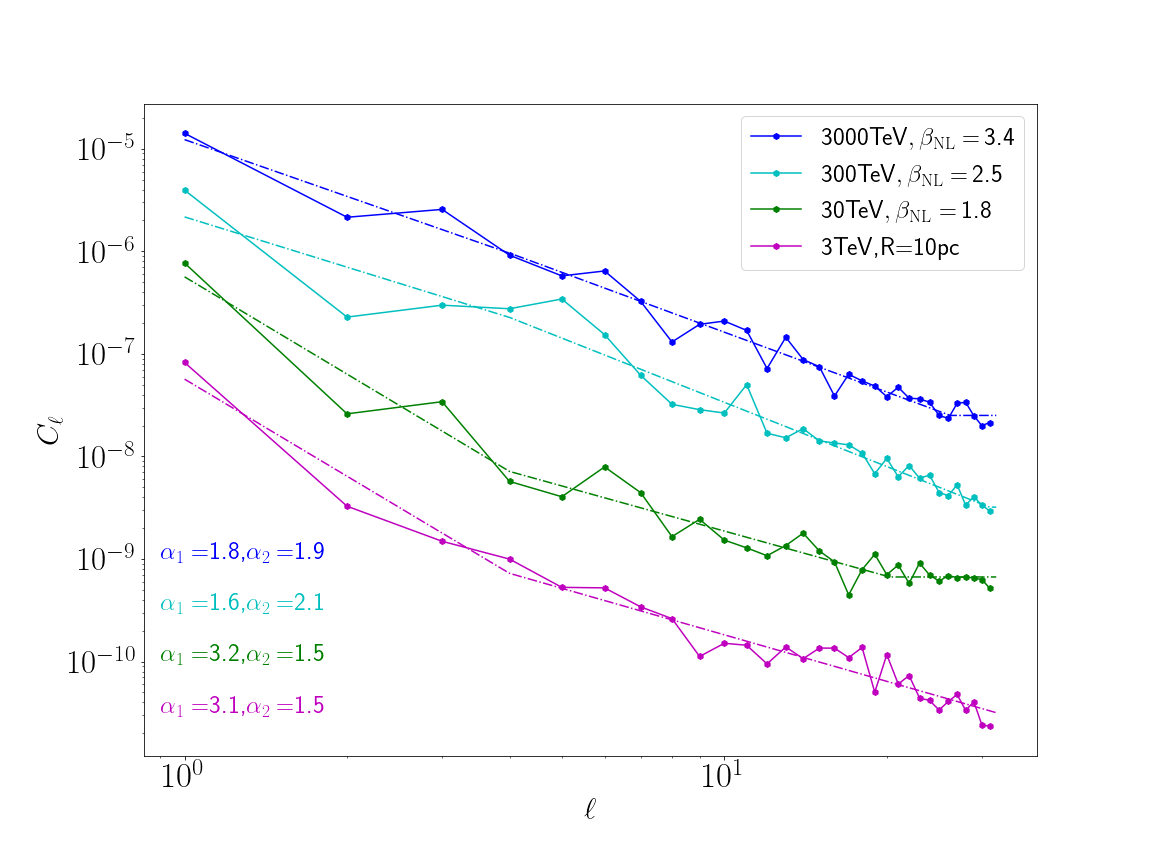}
  \caption{Angular power spectrum at different CR energies (3, 30, 300, 3000\,TeV). The three panels from top to bottom correspond to Location 1, Location 2, and Location 3, respectively. The keys in the upper-right corners of each panel show the CR energy and the corresponding $\beta_{\mathrm{NL}}$ from the equations (\ref{eq:beta_index}) and (\ref{eq:bkpowelaw}).
    The keys in the lower-left corner provide the two best-fit parameters from the broken power-law fit with $\ell_{\mathrm{break}}=4$.}
  \label{fig:diff_RER_spectrum_noise}
\end{figure}

\begin{figure}[t]
  \centering
    \includegraphics[width=\linewidth]{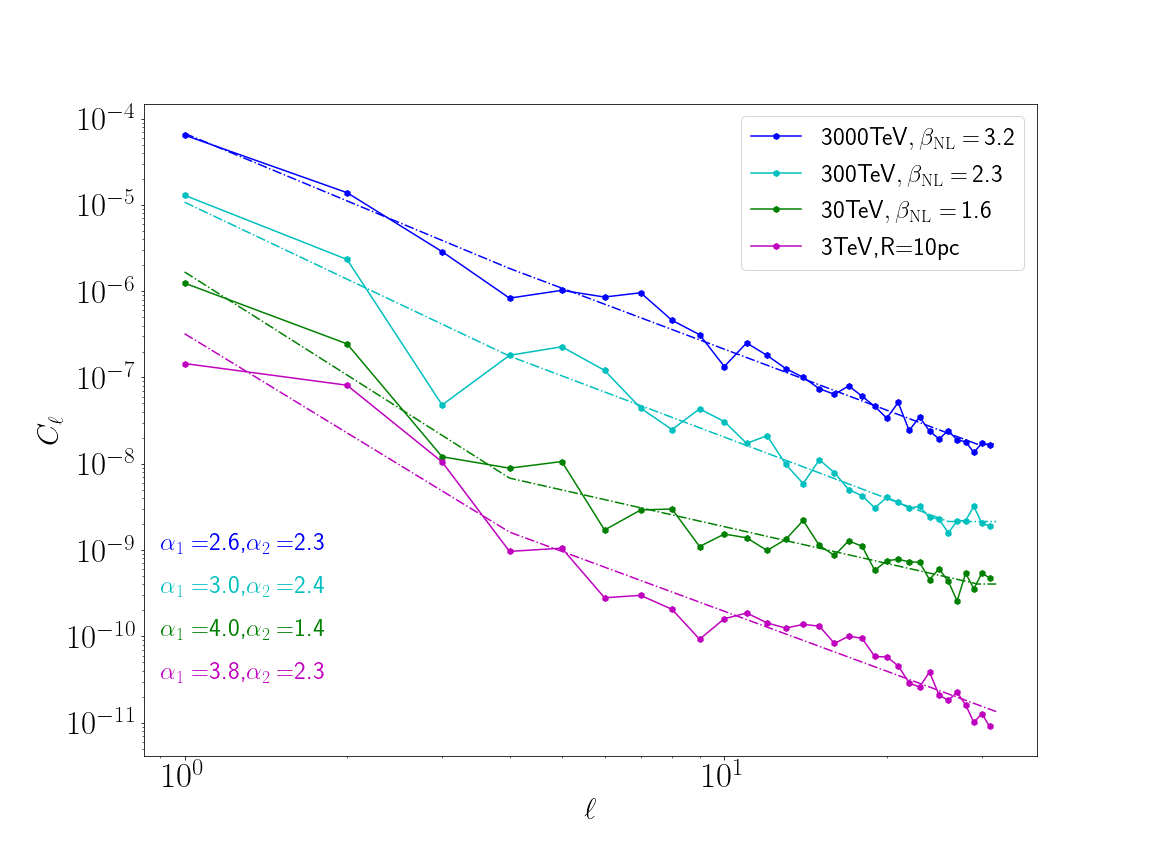}
    \includegraphics[width=\linewidth]{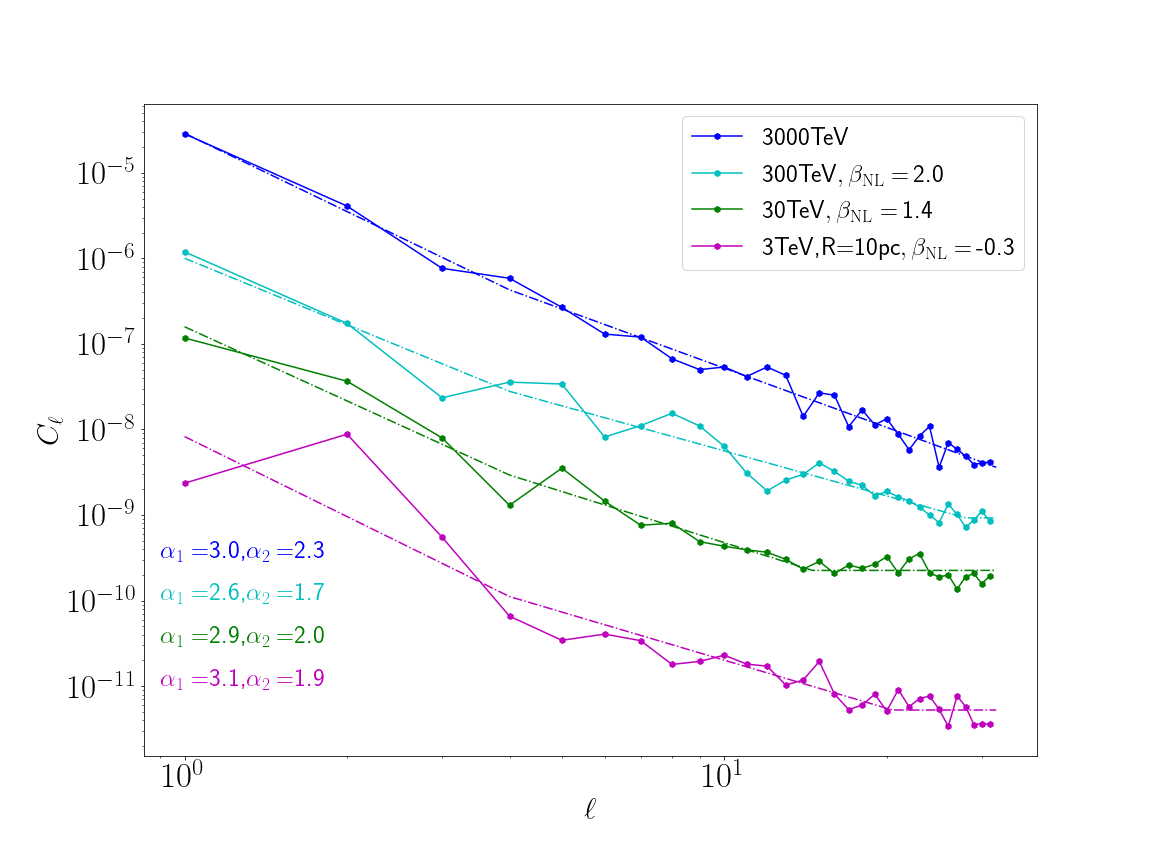}
  \caption{Same as in Figure~\ref{fig:diff_RER_spectrum_noise}, but for an observer at Location 4 (upper panel) and Location 5 (lower panel).
}
  \label{fig:diff_RER_spectrum_noise_2}
\end{figure}

\section{\label{Energy}Energy Dependence and Comparison with Observations}
In this section, we calculate the angular power spectrum for different CR energies.

The local gradient of the CR density around the Earth is an unknown parameter.
We first show below that the shape of the angular power spectrum  does not depend on the value of this parameter.  This value only changes the absolute normalization of the $C_{\ell}$ versus $\ell$ curve. 
In Figure~\ref{fig:grad}, we present different angular spectra for different values of ${\nabla n/n_0}$.
We calculate there spectra with 3$\,$TeV CR energy at Location 2 in the turbulence and change the CR density gradient in the simulations. The initial value of ${\nabla n/n_0}$ in the y-direction is set to 3.5$\,\mathrm{kpc}^{-1}$. We reduce three times this value by half down to 0.4375$\,\mathrm{kpc}^{-1}$.
It can be observed that the $C_{\ell}$ values of different multipoles can be approximately described in one same function of ${\nabla n/n_0}$ 
which is $C_{\ell}=M_{\ell}(\frac{\nabla n}{n_0})^k$.  
In our simulation results, $\mathrm{k}\sim2$. 
Considering the formula $|\vec{\delta }|\propto{|\vec{\nabla}n|}/{n_0}$ \citep{berezinskii1990cosmic}, 
this k value is reasonable.

We now study the impact of CR energy and observer location on the angular power spectrum.
In Figures~\ref{fig:diff_RER_spectrum_noise} and \ref{fig:diff_RER_spectrum_noise_2}, we present the anisotropy angular power spectrum at the five locations shown 
in Figure~\ref{fig:all_simulation_skymap} for different CR energies.
We use the following equation to fit the calculated results of the angular power spectrum, where noise is also included in the fitting process:

\begin{eqnarray}
  C_{\ell} = \begin{cases}
    A \left(\frac{\ell}{\ell_{\text{break}}}\right)^{\alpha_1}, & \text{if } \,\ell \le \ell_{\text{break}}, \\
    A \left(\frac{\ell}{\ell_{\text{break}}}\right)^{\alpha_2}, & \text{if } \,\ell_{\text{break}}<\ell \le \ell_{\text{eff}}, \\
    \mathcal{N}, & \text{if }\,\ell > \ell_{\text{eff}}
  \end{cases}
  \label{eq:bkpowelaw}
\end{eqnarray}  

Here we use $\ell_{\mathrm{break}}=4$.
We find that such a broken power-law provides a better fit than a simple power-law.
This is due to the fact that it fits the angular power spectrum at large and small $\ell$ separately. A likely physical justification for this break is that the formation mechanisms of the large-scale anisotropy (such as the dipole and quadrupole moments) differ from those of small-scale anisotropies.
Although the $\ell_{\mathrm{break}}$ parameter can a priori take any value, we find that a value greater than or equal to 4 is required to eliminate the influence of large-scale structures on the power-law fit of the angular power spectrum at higher $\ell$.

In Figure~\ref{fig:all_simulation_skymap},  the amplitudes of the anisotropy sky maps at the top three rows for Location 1, Location 2, and Location 3 are relatively symmetric, with the maximum values being almost the negatives of the minimum values. However, at Location 4 and Location 5, there is a clear asymmetry in the amplitudes. We find that the reason for this amplitude asymmetry is the sharp gradient of the mean free path of CRs around the observer's location. 
In the angular power spectrum at Location 4 and Location 5, for 3 TeV, the weight of the dipole is not as significant compared to the other three positions. In fact, at Location 5, the $C_{\ell=2}$  is greater than the $C_{\ell=1}$ for 3 TeV as can be seen in Figure~\ref{fig:diff_RER_spectrum_noise_2}.  The results of the sky-maps and angular power spectra at Location 4 and Location 5 imply that if there is a magnetic field environment in the Galaxy similar to the simulated magnetic field environment, it is possible to observe such anomalous anisotropies and the corresponding behavior in the angular power spectra.
At location 2, in the results with 3$\,$TeV and 30$\,$TeV, the overall structure of the anisotropy seems more axisymmetric along the direction of the dipole. At location 3, the axisymmetry of anisotropy is less pronounced, and small-scale anisotropy structures are more apparent compared to Location 2 for 3$\,$TeV and 30$\,$TeV. This difference in anisotropy is found to be a result of the variations in the turbulence level $\delta \mathrm{B}/\mathrm{B}$ at different observer positions. 
In the fits, the $\alpha_1$ values are highly dependent on the weight of large-scale structures
which is sensitive to the observer location in the turbulence.
Therefore, in Figure~\ref{fig:diff_RER_spectrum_noise} and Figure~\ref{fig:diff_RER_spectrum_noise_2},
the values of $\alpha_1$ for the angular power spectrum at each location differ significantly.
At the same location, the differences in $\alpha_2$ values at different energies are minimal and
 for most cases the values of $\alpha_2$ are approximately 2.
This suggests that the power-law fit of the smaller-scale anisotropy structures is not dependent on the CR energy at the same location.
At Location 2, the anomalous result of the fit for the 30 TeV angular power spectrum is due to the fact that the power of some multipoles structures, such as $\ell=$ 3 and 5, is approaching the noise level. 

\begin{figure*}
  \centering
    \includegraphics[width=0.49\linewidth]{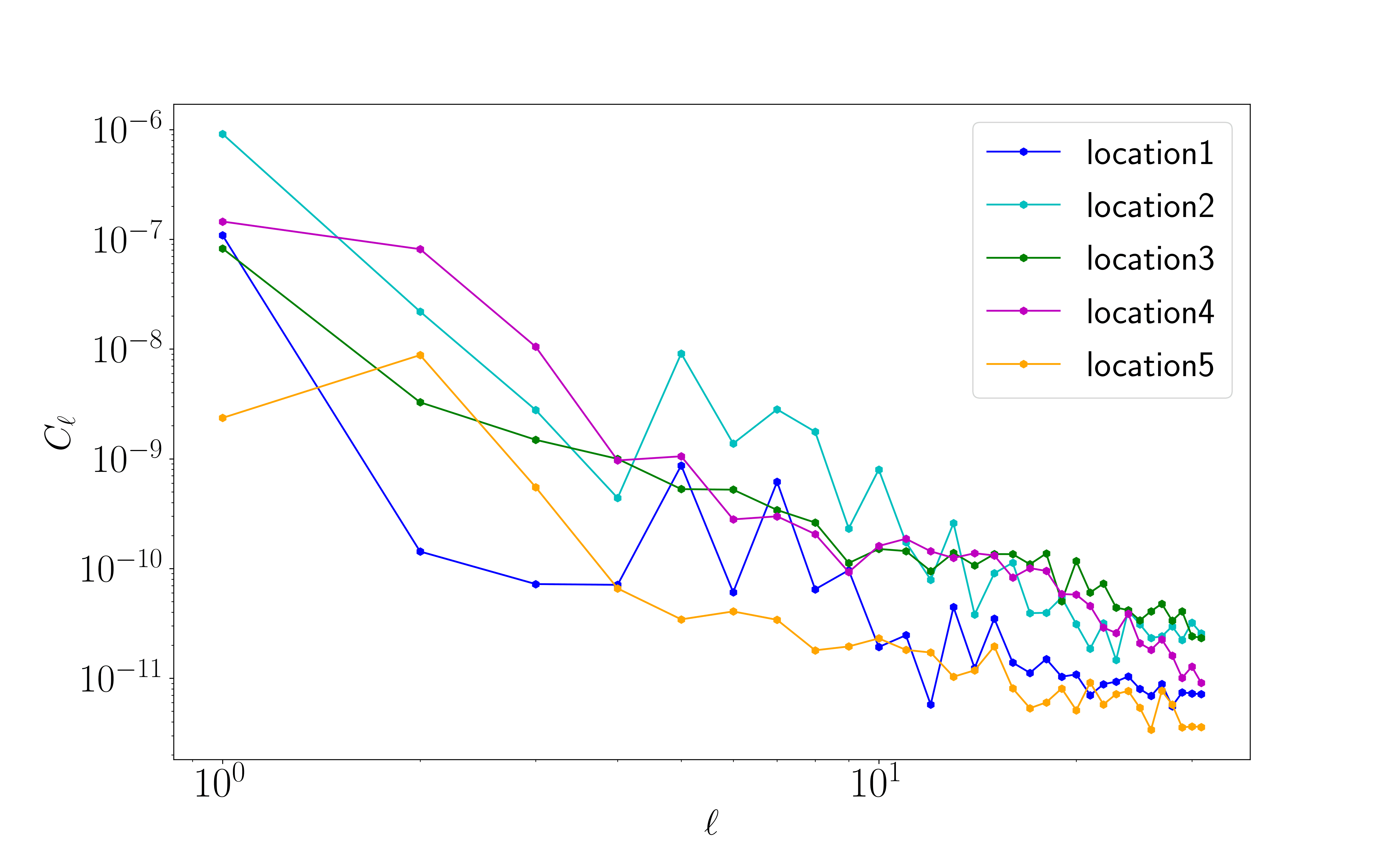}
    \includegraphics[width=0.49\linewidth]{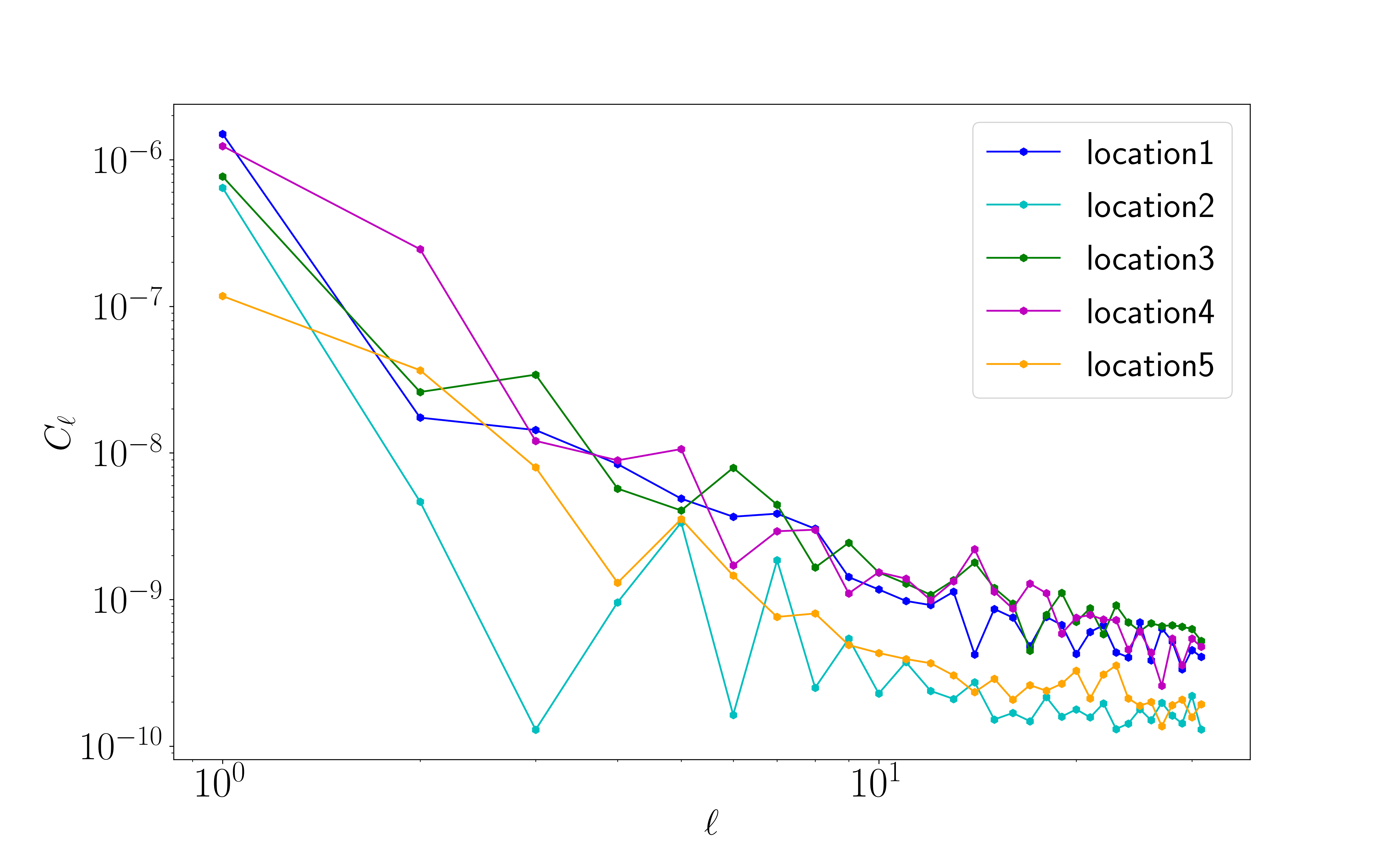}
    \includegraphics[width=0.49\linewidth]{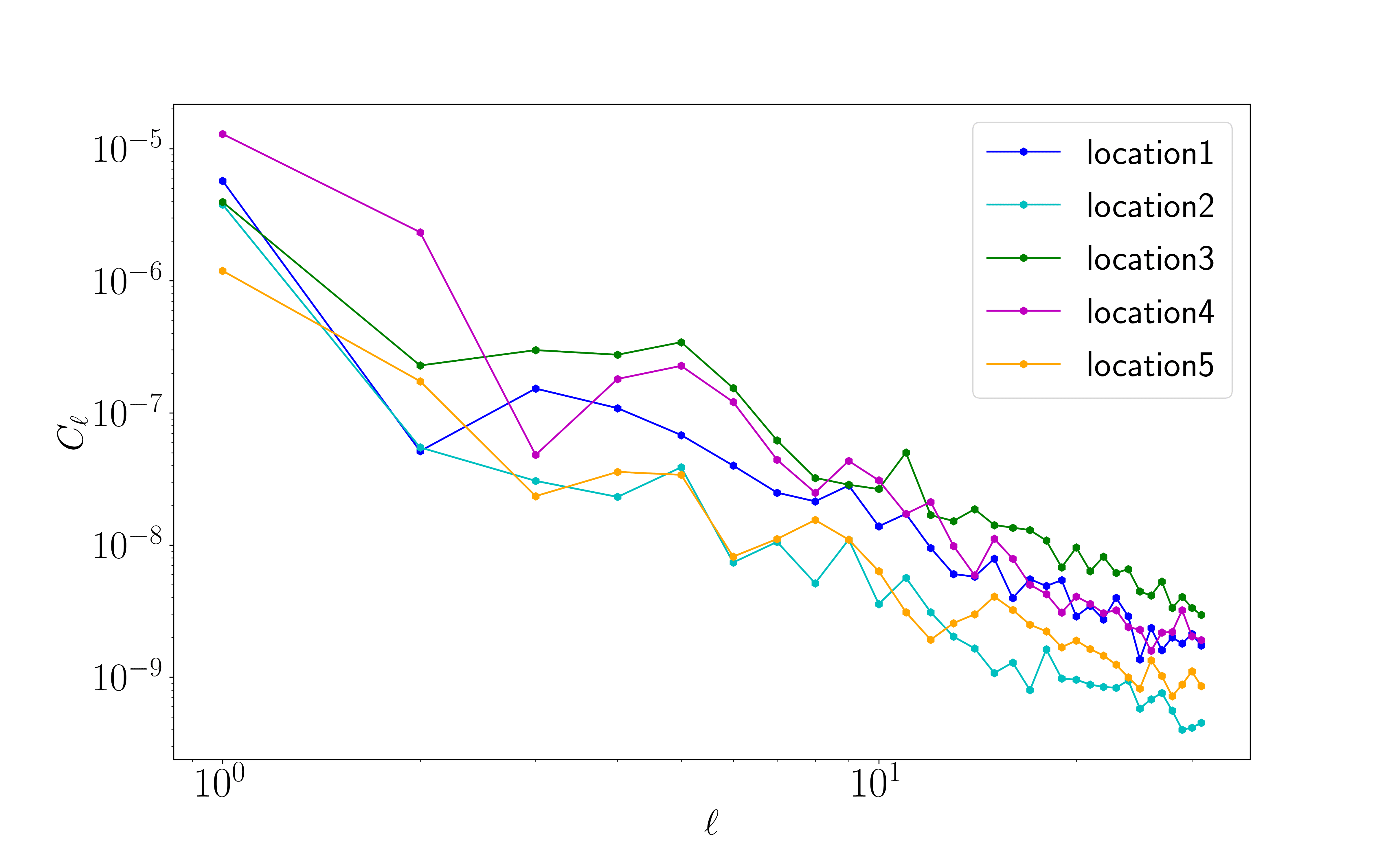}
    \includegraphics[width=0.49\linewidth]{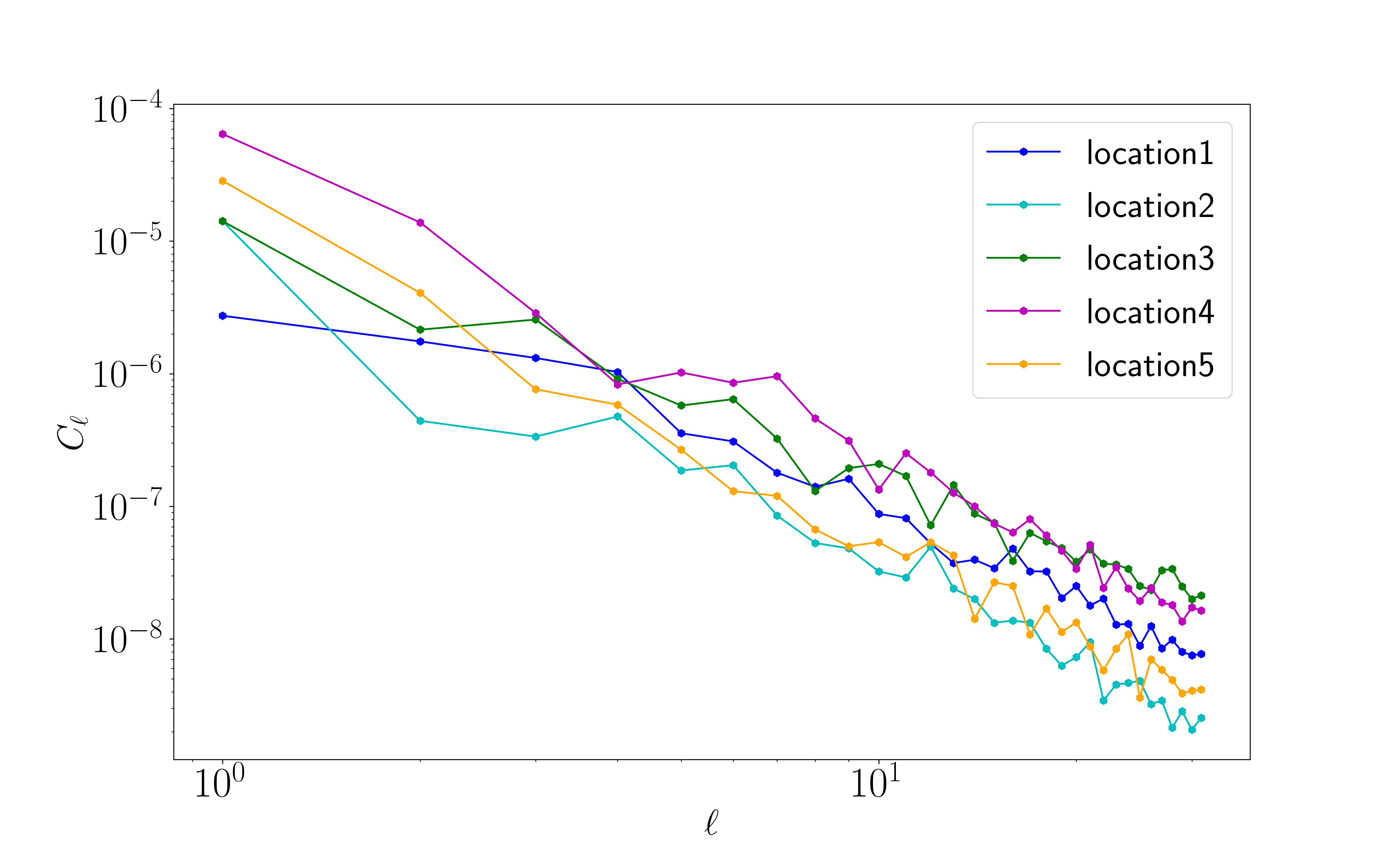}
  \caption{
Anisotropy angular power spectrum for the five observer locations (see the keys) at fixed given energies. The CR energies in the four panels are: 3$\,$TeV (upper left panel), 30$\,$TeV (upper right), 300$\,$TeV (lower left) and 3000$\,$TeV (lower right).}
  \label{fig:diff_location}
\end{figure*}


In Figure~\ref{fig:diff_location}, unlike Figures~\ref{fig:diff_RER_spectrum_noise} and~\ref{fig:diff_RER_spectrum_noise_2} where each panel corresponds to a different location, each panel shows the angular power spectra at all five locations in the simulation for different energies. We can clearly see the impact of location in the turbulent magnetic field on the angular power spectrum. At the same simulated CR energy, the amplitude of the power spectrum at different locations can vary within a range. 
The angular power spectrum of large-scale structures is clearly dependent on the observer's location. The formation of the large-scale anisotropy is closely related to the surrounding magnetic field environment, and in particular to the shape of the local magnetic flux tube containing the observer.
The slopes of the angular power spectra of small-scale structures at different locations are all similar. However, their amplitude at large $\ell$ could vary at different positions. The fit results in Figure~\ref{fig:diff_RER_spectrum_noise} and Figure~\ref{fig:diff_RER_spectrum_noise_2} show that the $\alpha_2$ value for each location are concentrated around 2. 

\begin{figure}
  \centering
    \includegraphics[width=\linewidth]{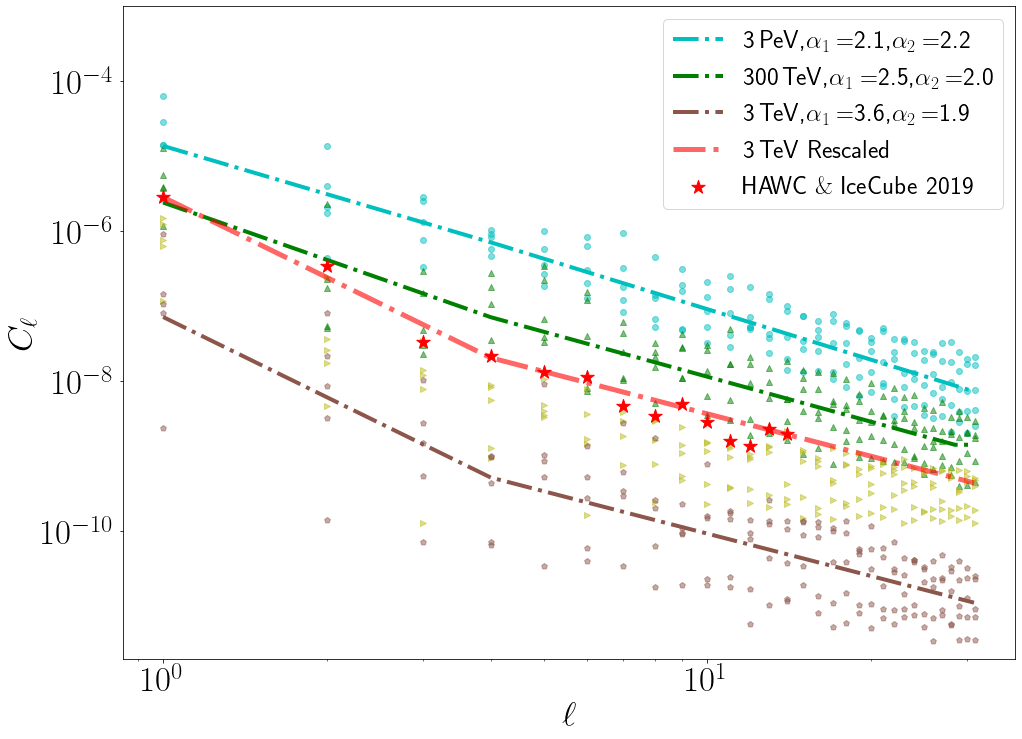}  
  \caption{Best fits of $C_{\ell}~\text{versus}~\ell$ for all five observer locations fitted together, using equation (\ref{eq:bkpowelaw}).  
  The top cyan-blue dashed-dotted line represents the fit at 3 PeV, 
  the middle green line is for 300 TeV, 
  and the bottom brown line is for 3 TeV. See the key for the corresponding values of $\alpha_1$ and $\alpha_2$.
  The individual points correspond to the $C_{\ell}$ values for each configuration as in Figure~\ref{fig:diff_location}. The light green dots correspond to the 30\,TeV results, although we do not fit them here. 
  The red stars represent the measurements at $\sim$ 10\,TeV in the combined HAWC and IceCube analysis from \cite{abeysekara2019all}. The red dashed-dotted line is the brown line rescaled by a factor 30.}
  \label{fig:broken_fit}
\end{figure}

In Figure~\ref{fig:broken_fit}, the cyan dashed-dotted line represents the fit of the anisotropy for all five locations at 3$\,$PeV, the green line is the fit for 300$\,$TeV, and the brown line is the fit for 3$\,$TeV.
In Figure~\ref{fig:all_simulation_skymap}, at all locations, the structures show sharper differences in the anisotropy sky maps in larger-scales when the energies in the simulation are relatively low.
As the energy in the simulation increases, the intermediate-scale structures in the anisotropy sky maps become more pronounced.
The overall structure of the anisotropy sky maps no longer exhibits a strong dipole moment structure compared with the results for lower energies.
Overall, the first stage of the angular power spectrum fit at $\ell<\ell_{\text{break}}$ tends to flatten, while the slope of the second stage at $\ell>\ell_{\text{break}}$ remains relatively constant, and the $\alpha_2$ value stays approximately equal to 2.
The red stars represent the results up to $\ell=14$ from the joint full-sky analysis at 10\,TeV median energy from HAWC and IceCube~\citep{abeysekara2019all}. These data points happen to be roughly a factor 30 above our brown line for 3\,TeV. However, as shown in Fig.~\ref{fig:grad}, the absolute normalization of the power spectrum depends on the unknown value of the local CR density gradient, which can therefore be rescaled. For this reason, we show with the red dashed-dotted line our rescaled 3$\,$TeV fit. It clearly provides a good fit to the data. This demonstrates that the measured power spectrum of the CR anisotropy at $\sim 10$\,TeV is compatible with theoretical expectations for Kolmogorov turbulence with strengths and coherence lengths relevant to the turbulence in the ISM.

\section{\label{Rotate}Harmonic coefficients $f_{\ell}^m$ with rotated Anisotropies}

\begin{figure*}[ht]
  \centering
    \includegraphics[width=0.49\linewidth]{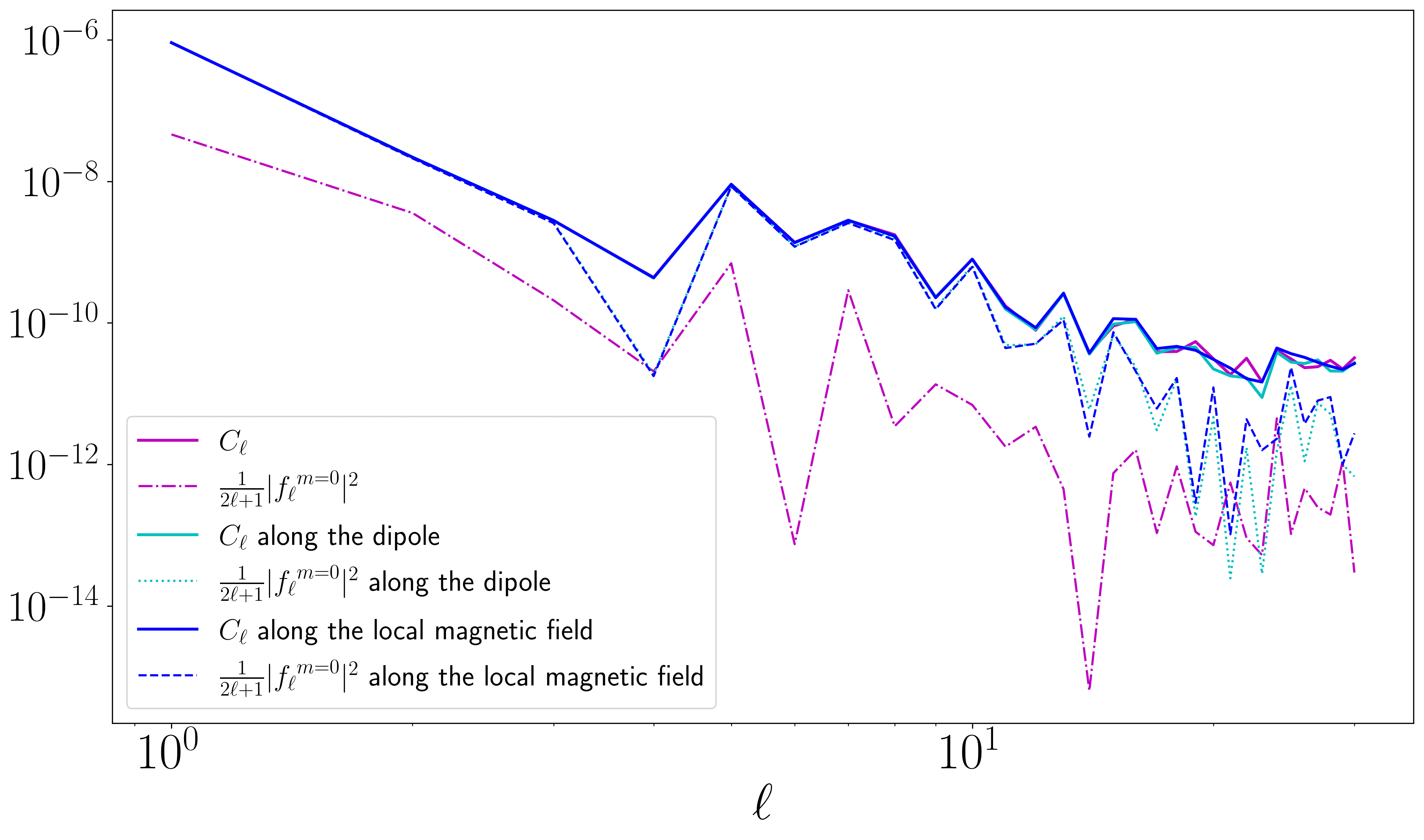}
    \includegraphics[width=0.49\linewidth]{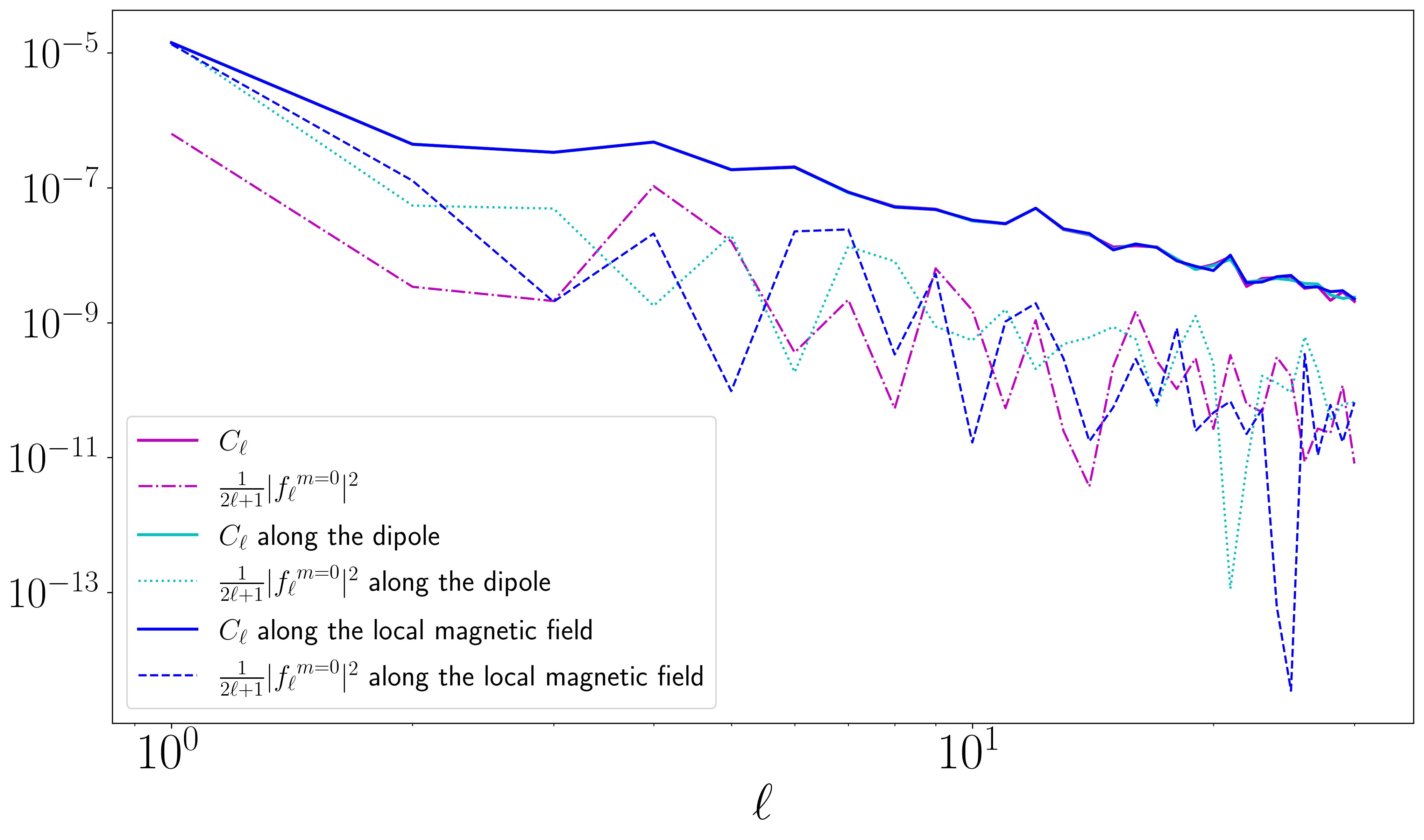}
    \includegraphics[width=0.49\linewidth]{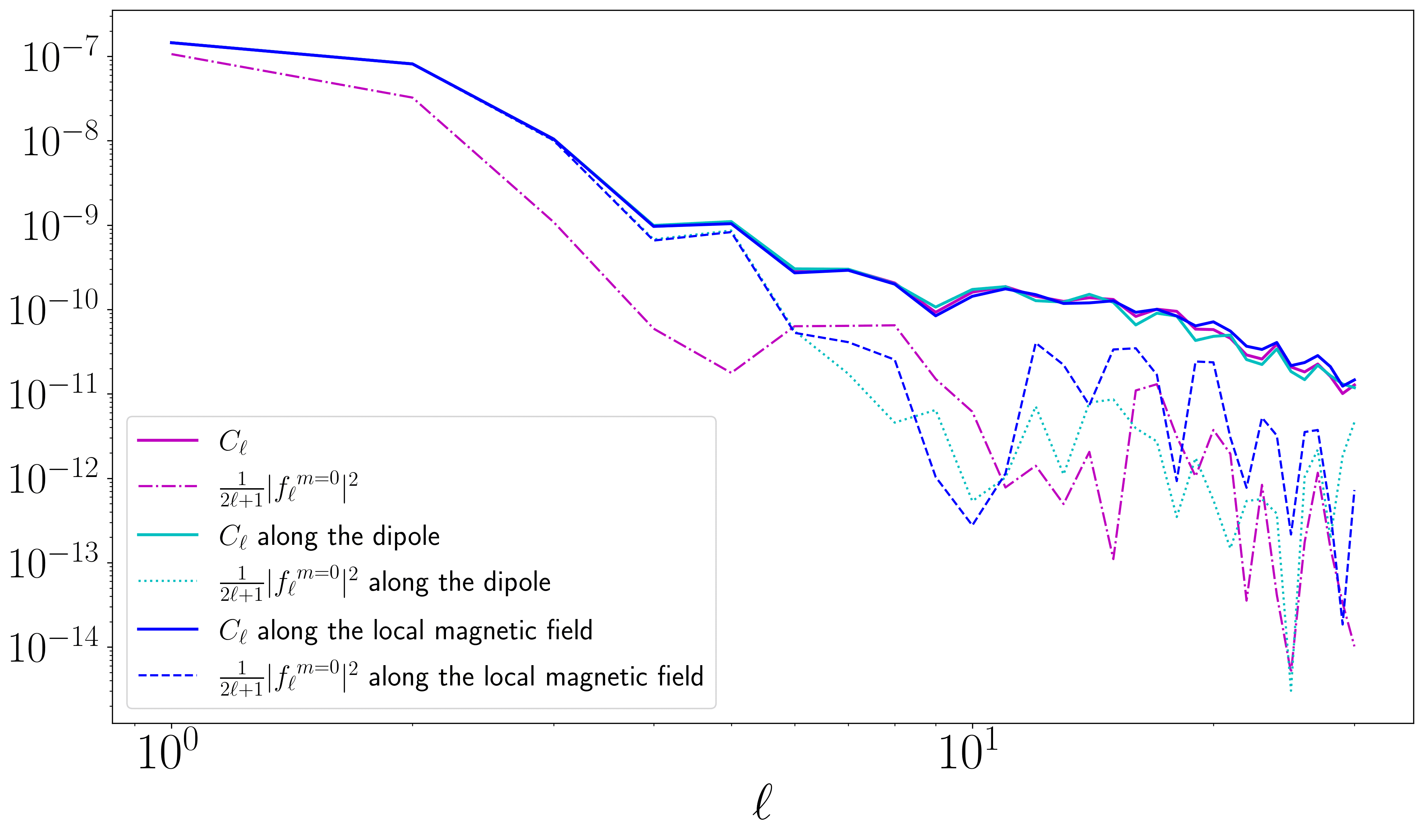}
    \includegraphics[width=0.49\linewidth]{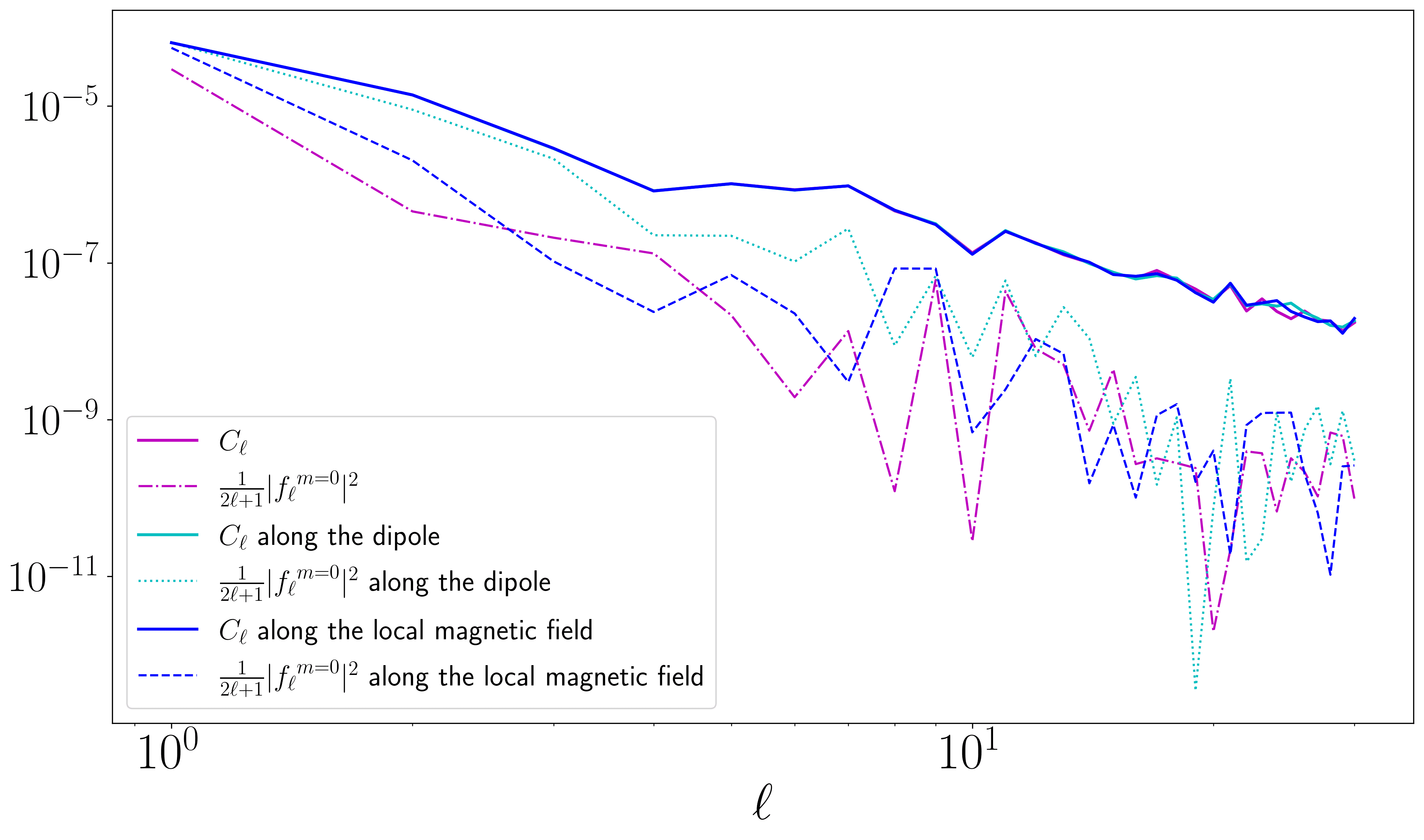}
  \caption{Angular power spectrum calculated along three different directions. The magenta lines correspond to the case where $\theta=0$ along the z-axis of the simulated box. The cyan (resp. blue) lines are for the case where $\theta$ is set to 0 along the local direction of the dipole anisotropy (resp. magnetic field line) at the observer location. The solid lines are for the values of $C_{\ell}$ and the other line types show the values of $|f_{\ell}^{m=0}|^2$. 
  The first row corresponds to Location 2, 
  and the second row to Location 4. 
  The first and second columns represent the energies of 3 TeV and 3 PeV, respectively.}
  \label{fig:Rotated_flm}
\end{figure*}

Combining the expansion coefficients $f_{\ell}^m$ into the angular power spectrum $C_{\ell}$ may result in a loss of information about the multipole structures of the anisotropy. Therefore, analyzing all the expansion coefficients $f_{\ell}^m$ is also crucial, providing greater insights into the local turbulent magnetic field structure.
In Figure~\ref{fig:Rotated_flm}, we select one location with a symmetric anisotropy amplitude (Location 2, top row) and an asymmetric one (Location 4, bottom row).  
In the left column, we show the calculations of the angular power spectra at 3$\,$TeV, and in the right column, at 3$\,$PeV.
The values of the coefficients $|f_{\ell}^{m=0}|$ represent the weights of the anisotropy multipole structures 
which are axisymmetric along the direction $\theta=0$.
Each panel in Figure~\ref{fig:Rotated_flm} shows the values of $C_{\ell}$ calculated along three directions (z-direction of the turbulence, the observed dipole direction and the local magnetic field line direction) and the values of $|f_{\ell}^{m=0}|^2$ with the same color code at each location and CR energy.
As expected, analyzing along three different directions has no significant impact on the $C_{\ell}$ values. 
However, the values of $|f_{\ell}^{m=0}|^2$ are larger when calculated along the dipole direction (dotted lines) and the local magnetic field line direction (dashed lines), and the $|f_{\ell}^{m=0}|^2$ curves are then also closer to the $C_{\ell}$ curves. This implies that these multipole structures have a preferred direction which is consistent with the dipole direction or the local magnetic field line direction. In the TeV-PeV range, these two directions are almost the same or opposite. This is because CRs undergo helical motion along magnetic field lines, making the gyro-symmetric anisotropy structures more pronounced. For example, \cite{giacinti2017large} focused on the study of these gyrotropic anisotropy structures.
As can be seen in the left column of Figure~\ref{fig:Rotated_flm}, at 3\,TeV, the values of $|f_{\ell}^{m=0}|^2$ move close to those of $C_{\ell}$ up to $\ell \approx 10$ when they are calculated along the dipole or magnetic field line directions. In contrast, at 3\,PeV (right column), this is only true for the first few multipoles, up to $\ell \approx 4$. For larger values of $\ell$, the levels of the $|f_{\ell}^{m=0}|^2$ curves do not substantially change when they are calculated along the dipole or magnetic field line directions. This implies that such small-scale anisotropies are distributed more randomly on the sky, and do not have such a preferred direction.

\bigskip
\section{\label{Discuss&Conclusion}Discussion and Conclusions}

In this study, we have presented the first numerical simulations of the CR anisotropy down to 3\,TeV energies. We have propagated individual CRs in 3D turbulent magnetic fields with a realistic value of their coherence length $L_{\rm c}\sim 10$\,pc. Previous numerical studies were all restricted to $\gtrsim$ PeV CR energies, which is substantially larger than the energies at which the angular power spectrum of the anisotropy is measured by HAWC and IceCube~\citep{abeysekara2019all}. As summarized below, we find that the shape of the angular power spectrum depends on the CR energy. This implies that one needs to use numerical simulations down to TeV energies in order to compare theoretical predictions to measurements. In particular, previous simulations using a ratio of the CR gyroradius to the turbulence coherence length equal to 0.1 provide a somewhat different prediction than our simulations made at a few TeV, where this ratio is as low as $\sim 10^{-4}$. The fact that the power spectrum depends on CR energy is not unexpected. Indeed, there is less power in the modes that scatter TeV CRs than in those that scatter PeV CRs. Moreover, the turbulent magnetic field looks more ordered on the scale of a TeV CR mean free path, than it is on the scale a PeV CR mean free path. 

Simulations at TeV energies are however computationally expensive. Indeed, for a fixed backtracking sphere radius, they require to propagate a larger number of CRs in order to probe the high-order multipoles, because of the smaller CR mean free path. Moreover, low-energy CRs have a smaller gyroradius, which further increases the computing time. Usually, it is preferable to propagate CRs over distances much larger than the coherence length of the turbulence, to remove any influence of those scales on the results. However, we have shown above that it is possible to reduce the backtracking sphere radius down to scales smaller than the coherence length (but still bigger than a CR mean free path), without affecting the shape of the power spectrum. This provides a new method for reducing the computing time and the noise level at large $\ell$ in simulations at TeV energies. We find that using such smaller backtracking sphere radii only results in small changes in the absolute normalization of the amplitude of the anisotropy, without distorting the shape and power spectrum slope of its small-scale anisotropies.

For clarity, we have restricted our study to the case of Kolmogorov turbulence. The Boron-to-Carbon ratio measurements from AMS-02 experiment are indeed compatible with Kolmogorov turbulence up to $\sim 2$\,TV CR rigidity~\citep{AMS_BoC}. Whether such a turbulence type is also relevant for the scattering of $\gtrsim 10$\,TeV CRs in the ISM still remains uncertain. Other types of turbulence will be studied in future works.

Our main findings can be summarized as follows:

\begin{itemize}
    \item The amplitude of the local CR density gradient, which is poorly known but drives the formation of the CR anisotropy, does not affect the angular structure of the anisotropy. It only affects the absolute normalization of the anisotropy amplitude, but not the shape of its angular power spectrum. This confirms that the shape of the anisotropy contains the signatures of the local turbulence properties, and is not affected by the unknown history and locations of recent CR sources around the Earth. The power spectrum can therefore be safely used to study the properties of the interstellar turbulence.
    
    \item In general, the distribution becomes more and more gyrotropic with decreasing CR energies. At $\sim 1 - 10$\,TeV energies, the anisotropy aligns well with the (random) local direction of the magnetic field in the turbulence around each observer ---and not with the direction of the imposed CR density gradient. We also find that the relative amplitude of the small-scale anisotropies to that to the large-scale anisotropy depends on the local level of turbulence on the resonant scales that scatter the CRs at the observer location. In regions where the apparent \lq\lq ordered\rq\rq\/ magnetic field, due to the modes with wavelengths much larger than the CR gyroradius, is stronger the anisotropy is more gyrotropic. Turbulence levels in those regions appear to be smaller. See, for example, the second panel in the first column of Figure~\ref{fig:all_simulation_skymap} for a case of gyrotropy, and the panel directly below for a case of angyrotropy.
    
    \item The angular power spectrum can fluctuate non-negligibly between different locations of the observer. This might therefore slightly complicate the interpretation of the power spectrum measured at Earth, because it may then not be equal to the ensemble-average over many observer locations. However, our results show that these fluctuations from one observer location to another are most pronounced at small values of $\ell$, and that their impact on the slope of the power spectrum at larger $\ell$ is negligible. It is therefore possible to draw useful, generic conclusions on our local turbulence, and on CR scattering in the ISM, from the power spectrum measured at Earth. The fact that the observer location mostly affects the power spectrum at small $\ell$ is compatible with the intuition that the large-scale structure of the anisotropy depends on the random shape of the local interstellar magnetic flux tube, while the shapes of the smaller-scale anisotropies should not be strongly affected by it.
    
    \item For Kolmogorov turbulence, the power spectrum displays a noticeable dependence on CR energy in the TeV--PeV range, which had not been predicted before. At $\gtrsim$\,PeV energies, the spectrum approximately follows a power law with a $\approx -2$ slope: $C_\ell \propto \ell^{-2}$. At lower energies, a broken power law provides a better fit. For multipoles with $\ell \lesssim 4$, $C_\ell$ can be fitted with a power law with a steeper slope, and which becomes softer with decreasing energy ($C_\ell \propto \ell^{-3 ... -4}$ at a few TeV). In contrast, the power spectrum at $\ell \gtrsim 4$ can be fitted with a power law with a $\approx -2$ slope, as for PeV energies. Our dependence of $C_\ell$ on $\ell$ at large values of $\ell$ is slightly harder than the $C_\ell \propto \ell^{-3}$ dependence predicted by \cite{AhlersPRL2014}, and the $C_\ell \propto \ell^{-2.7 ... -3.3}$ dependence of \cite{ahlers2015small}. Our results at low CR energies do however provide a good fit of the newer combined HAWC and IceCube power spectrum~\citep{abeysekara2019all}. See the red dashed-dotted line and the red stars in our Figure~\ref{fig:broken_fit}.
    
    \item At some observer locations, the dipole amplitude at $\sim$\,TeV energy is weak compared with the quadrupole amplitude. The large-scale anisotropy then displays an asymmetric shape, where the amplitude of its maximum is substantially larger or smaller than the absolute value of the amplitude of its minimum. This happens in regions of the turbulence where the CR mean free path varies sharply in the surroundings of the observer. This does not seem to be a rare occurrence in synthetic Kolmogorov turbulence. We find two such locations in our simulations: Location 4 and Location 5, see the last two rows in Figure~\ref{fig:all_simulation_skymap}. Whether this is a frequent or a rare occurrence in the interstellar turbulence is  unknown. In any case, the combined large-scale ansiotropy measured by HAWC and IceCube~\citep{abeysekara2019all} does not seem to have such an asymmetric shape. Therefore, the Earth is apparently not located in such an environment.
    
    \item At $\sim$ TeV energies, we find that most of the power in the large- and medium-scale anisotropies ($\ell \lesssim 10$) is contained in the gyrotropic anisotropy, which is aligned with the direction of local magnetic field lines around the observer. This direction is almost the same as the direction of the dipole itself. At $\sim$ PeV energies, this is true only for modes with $\ell \lesssim 4$. In contrast, small-scale anisotropies with $\ell \gtrsim 10$, are found to be more randomly distributed on the sky. This can be seen in the two examples studied in Figure~\ref{fig:Rotated_flm}.
\end{itemize}

\section*{Acknowledgments}
This work is supported by the National Natural Science Foundation of China under Grants Nos. 12350610239, and 12393853.

\appendix

\section{Spherical Harmonics transform algorithm}
\label{sec:appendixA}

\begin{figure*}[ht]
  \centering
    \includegraphics[width=0.49\linewidth]{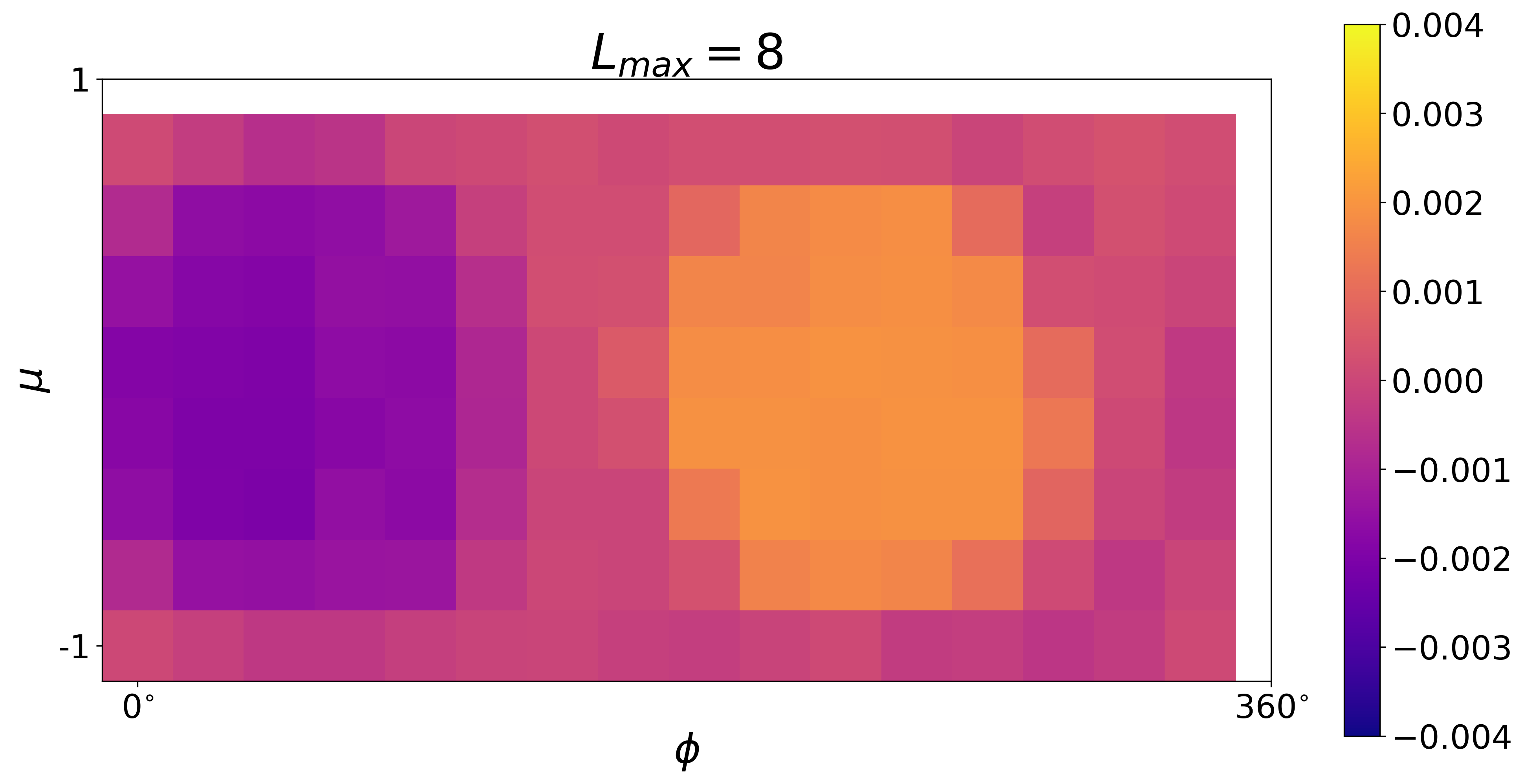}
    \includegraphics[width=0.49\linewidth]{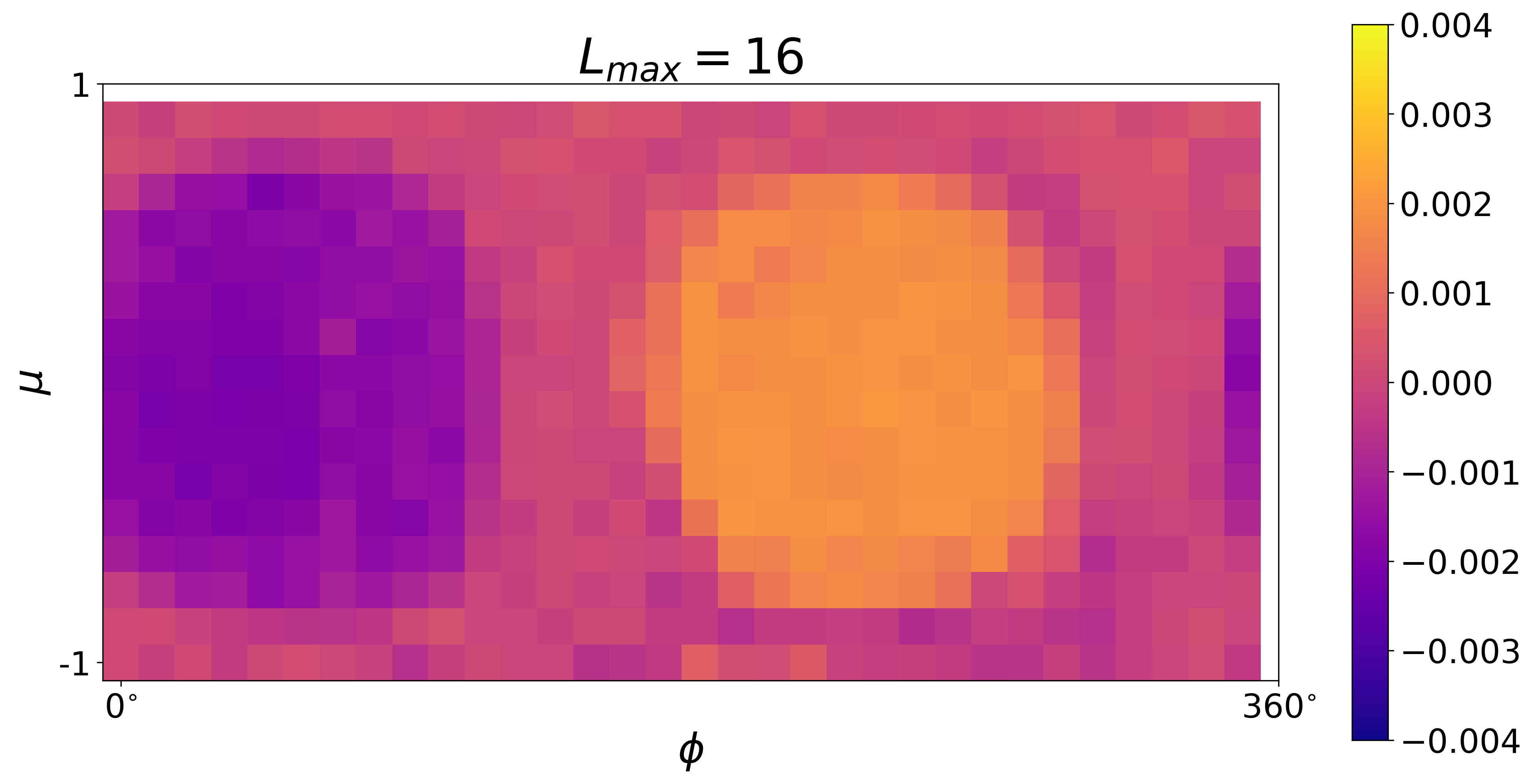}
    \includegraphics[width=0.49\linewidth]{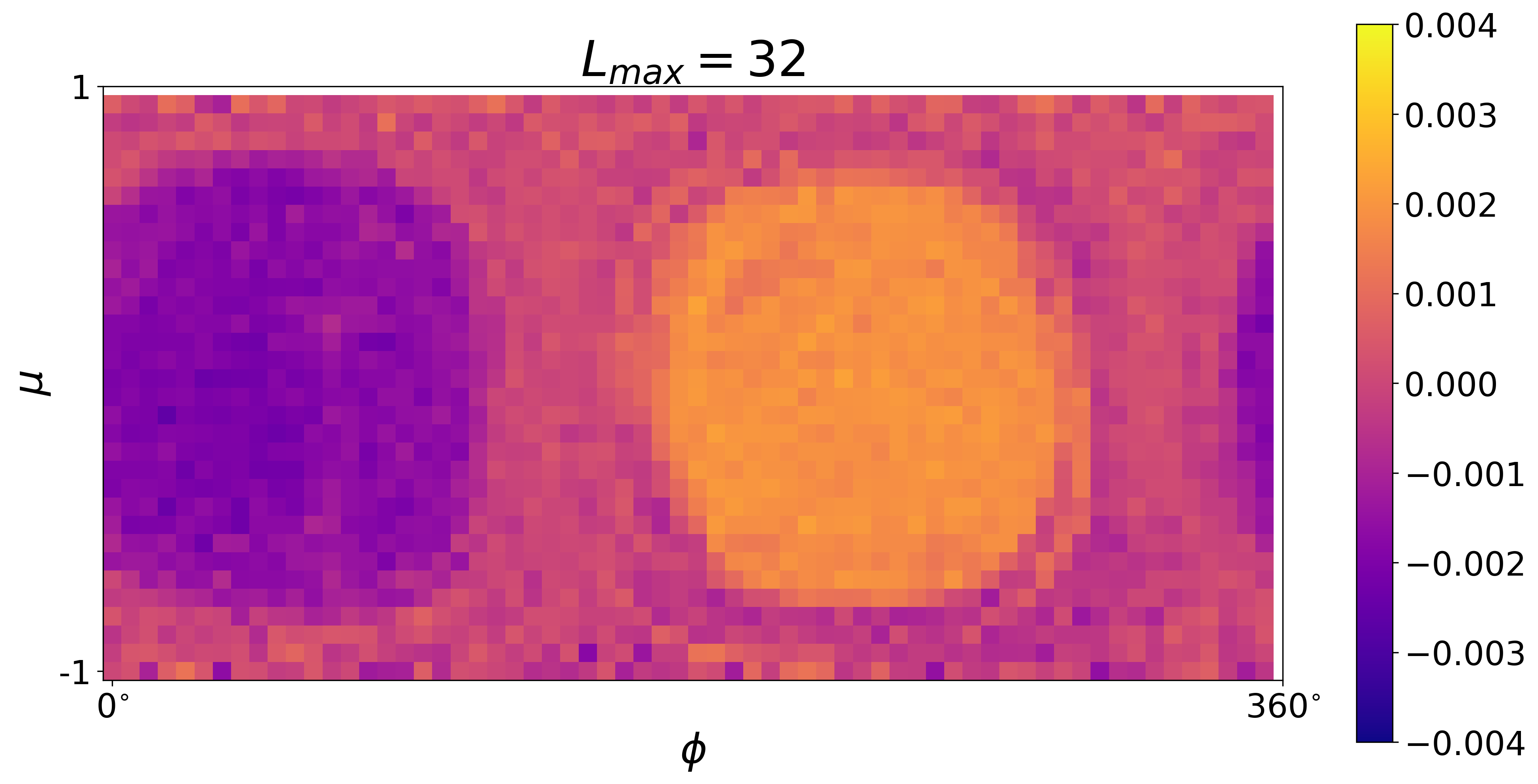}
    \includegraphics[width=0.49\linewidth]{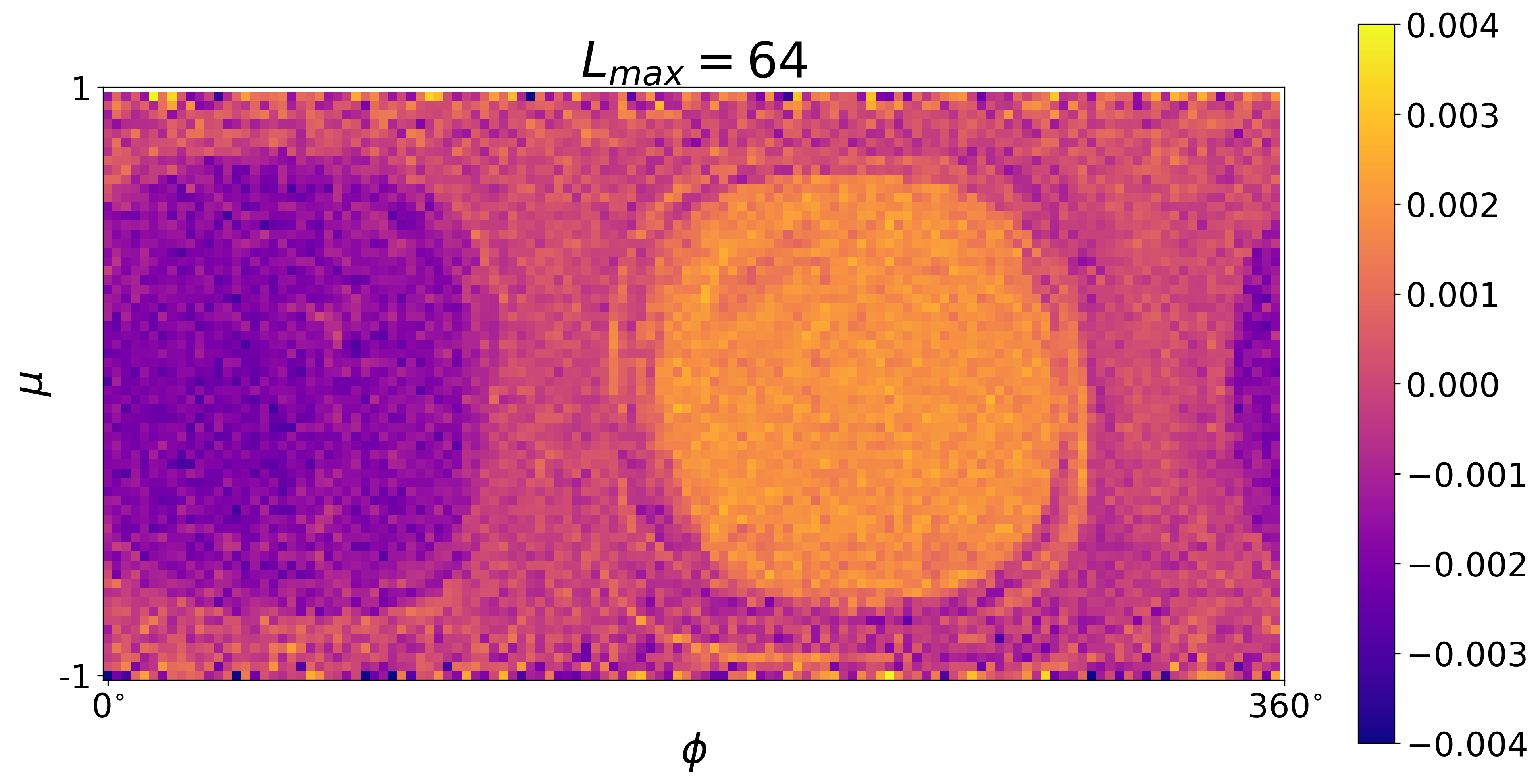}
    \includegraphics[width=0.49\linewidth]{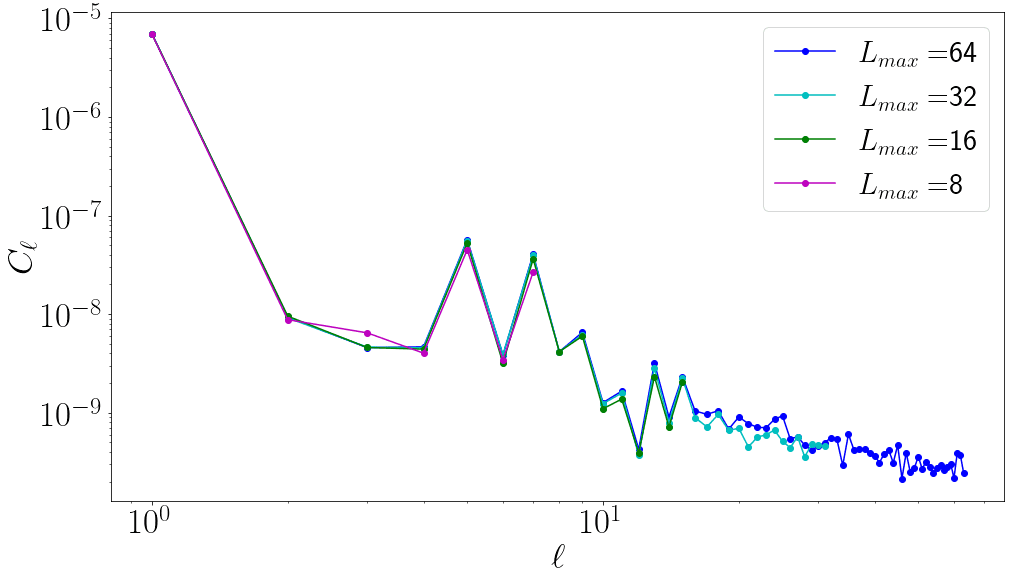}
  \caption{Comparison of spherical harmonic transforms for different choices of $L_{\rm max}$ for the case of 3 TeV at the Location 1 in Fig. \ref{fig:all_simulation_skymap}.}
  \label{fig:L_max_changesL_max_changes}
\end{figure*}

The spherical harmonics are defined as
\begin{equation}
Y_{\ell}^m(\theta,\varphi) = N_\ell^m
P_\ell^{m}(\cos\theta)e^{im\varphi},\qquad -\ell\leq m \leq \ell
\end{equation}
where $P_\ell^{m}(\cos\theta)$ are the associated Legendre polynomials including the Condon-Shortley
phase. Thus $P_\ell^{-m}(x) = (-1)^m \frac{(\ell-m)!}{(\ell+m)!}P_\ell^m(x)$, and by association $Y_\ell^{m*} = (-1)^mY_\ell^{-m}$. We choose the normalisation commonly used in quantum mechanics
\begin{equation}
N_{\ell}^m =  \sqrt{\frac{2\ell+1}{4\pi}\frac{(\ell-m)!}{(\ell+m)!}} 
\end{equation}
which results in 
\begin{equation}
\int \diff\Omega \, \left|Y_\ell^{m}\right|^2 = 1\,.
\end{equation}

Let $f(\theta,\varphi)$ be a known distribution on the sphere, in our case, the distribution of CRs at a fixed energy.
We seek the expansion coefficients $f_\ell^m$, such that
\begin{equation}
f(\theta,\varphi) = \sum_{\ell=0}^\infty\sum_{m=-\ell}^\ell f_\ell^m
Y_\ell^{m}(\theta,\varphi) \,.
\end{equation}
Since $f$ is real, $f_\ell^{m*} = (-1)^mf_\ell^{-m} $. 

From the orthonormality of the spherical harmonics, it follows that
\begin{equation}
\int  d \Omega |f|^2 
= \sum_{\ell=0}^\infty\sum_{m=-\ell}^\ell |f_\ell^m|^2 
=\sum_{\ell=0}^\infty (2 \ell +1) C_\ell
\end{equation}
where
\begin{equation}
C_\ell = \frac{1}{2 \ell +1 }\sum_{m=-\ell}^{\ell} \left|f_\ell^{m}\right|^2
\end{equation}
is the mean power per harmonics of order $\ell$.

Let us assume the expansion is truncated at a finite order $\ell = L$.
Changing the order of summation, one finds
\eqb
f (\theta,\varphi)&=& \sum_{\ell=0}^L \sum_{m=-\ell}^\ell f_\ell^m
Y_\ell^{m} \nonumber\\
&=& \sum_{m=-L}^L \left[\sum_{\ell=|m|}^L f_\ell^m
\sqrt{\frac{2\ell+1}{4\pi}\frac{(\ell-m)!}{(\ell+m)!}} 
P_\ell^{m}(\mu)\right]e^{im\varphi} \nonumber\\
&\equiv& \sum_{m=-L}^L  G_m(\mu) e^{im\varphi}\,,
\label{Eq:sumG}
\eqe
where $\mu=\cos \theta$. Note that
\[ G_m(\mu) = G_{-m}^*= 
\sum_{\ell=|m|}^L f_\ell^m
\sqrt{\frac{2\ell+1}{4\pi}\frac{(\ell-m)!}{(\ell+m)!}} 
P_\ell^{m}(\mu)\, . \]

Discretising $\varphi$ such that there are $N=2L$ uniformly spaced points between $0$ and $2 \pi$, i.e. $\varphi_n=2\pi n / N$ for $n=0,1,...,N-1$, it follows that for any value of $\mu$ (or $\theta$), the summation, eq (\ref{Eq:sumG}), takes the form of a discrete Fourier series
\eqbn
f (\theta,\varphi_n) &\equiv& f_n 
= \sum_{m=-L}^{L}  G_m(\mu) e^{2\pi i m n/N}\,.
\eqen 
whose inverse is
$$G_m(\mu) = \frac{1}{2\pi} \int _0^{2\pi} e^{-i m \varphi} f d\varphi=
\frac{1}{N} \sum_{n=0}^{N-1}  f_n e^{-2\pi i m n/N}\,.
$$

Finally, noting that $G_{-m} = G_{N-m}$, we see that this is exactly the inverse FFT algorithm used by \citet{NumericalRecipes}.
Thus $G_{m}$ can be found efficiently for any value of $\mu$. 

To find $f_\ell^m$, we use the orthogonality of the Associated Legendre polynomials 
\begin{equation*}
\int_{-1}^1 \diff\mu \, P_\ell^{m} P_{\ell'}^{m} = 
 \frac{2}{2\ell+1}\frac{(l+m)!}{(l-m)!}\delta_{\ell\ell'}
 \end{equation*}
which gives
\eqb
f_\ell^m = \sqrt{\pi(2\ell +1)\frac{(l-m)!}{(l+m)!}} \int G_{m}(\mu) P_{\ell}^m(\mu) d \mu  \,.
\eqe

We carry out the final integration over $\mu$ using Gauss-Legendre quadrature \citep[see][chapter 4.5]{NumericalRecipes}.

Examples of the projection of the sky map onto the Gauss-Legendre quadrature abscissa, and the associated numerically evaluated $C_\ell$ are shown in Figure \ref{fig:L_max_changesL_max_changes}, for 3\,TeV, $R=10$\,pc, 1 million CRs, and 4 different values of $L_{\max}$. It can be seen from this Figure that the shape of the anisotropy and its power spectrum do not change noticeably with $L_{\max}$ for these parameters, as long as $L_{\max} < 64$. This demonstrates that at 3\,TeV, and for $R=10$\,pc, backtracking 1 million CRs is sufficient to calculate numerically $C_\ell$ up to $\ell = 32$ without being affected by the noise.

\bibliography{Angular}{}
\bibliographystyle{aasjournal}



\end{document}